\documentclass[pdflatex,sn-mathphys]{sn-jnl}

\jyear{2021}%

\theoremstyle{thmstyleone}%
\theoremstyle{thmstyletwo}%

\theoremstyle{thmstylethree}%

\usepackage{cleveref}

\definecolor{red}{rgb}{0,0,0}

\begin{document}

\title[A Comparison of Parameter Estimation Methods for Shared Frailty Models]{A Comparison of Parameter Estimation Methods for Shared Frailty Models}








\author[1,2]{\fnm{Tingxuan} \sur{Wu}}

\author[3]{\fnm{Cindy} \sur{Feng}}
%

\author*[1]{\fnm{Longhai} \sur{Li}}\email{longhai@math.usask.ca}

\affil*[1]{\orgdiv{Department of Mathematics and Statistics}, \orgname{University of Saskatchewan}, \orgaddress{\street{106 Wiggins Rd}, \city{Saskatoon}, \postcode{S7N 5E6}, \state{SK}, \country{Canada}}}

\affil[2]{\orgdiv{School of Public Health}, \orgname{University of Saskatchewan}, \orgaddress{\street{104 Clinic Place}, \city{Saskatoon}, \postcode{S7N 5E5}, \state{SK}, \country{Canada}}}

\affil[3]{\orgdiv{Department of Community Health and Epidemiology}, \orgname{Dalhousie University}, \orgaddress{\street{5790 University Ave.}, \city{Halifax}, \postcode{B3H 1V7}, \state{NS}, \country{Canada}}}


\abstract{\textcolor{red}{This paper compares six different parameter estimation methods for shared frailty models via a series of simulation studies. A shared frailty model is a survival model that incorporates a random effect term, where the frailties are common or shared among individuals within specific groups. Several parameter estimation methods are available for fitting shared frailty models, such as penalized partial likelihood (PPL), expectation-maximization (EM), pseudo full likelihood (PFL), hierarchical likelihood (HL), maximum marginal likelihood (MML), and maximization penalized likelihood (MPL) algorithms. These estimation methods are implemented in various R packages, providing researchers with various options for analyzing clustered survival data using shared frailty models. However, there is a limited amount of research comparing the performance of these parameter estimation methods for fitting shared frailty models. Consequently, it can be challenging for users to determine the most appropriate method for analyzing clustered survival data. To address this gap, this paper aims to conduct a series of simulation studies to compare the performance of different parameter estimation methods implemented in R packages. We will evaluate several key aspects, including parameter estimation, bias and variance of the parameter estimates, rate of convergence, and computational time required by each package. Through this systematic evaluation, our goal is to provide a comprehensive understanding of the advantages and limitations associated with each estimation method.}}




\maketitle
\newcommand{\SE}{\mbox{SE}}
\section{Introduction}\label{sec1}

In survival analysis, conventional Cox proportional hazards models\cite{bib} and accelerated failure time models \cite{LIND.Y.1998Aftm}  assume that subjects are independent of one another. However, many research problems involve data with a multilevel structure, such as biomedical data or data pertaining to genetically related individuals, which exhibit correlation \cite{GovindarajuluUshaS.2011FmAt, BalanTheodorAdrian2019Nhau}. The hazard of the event differs from one cluster to another induced by unobserved cluster-level factors \cite{BalanTheodorAdrian2019Nhau, CollettD.1952-author2015Msdi}. Random effects can be incorporated into conventional survival models to account for cluster-level heterogeneity \cite{VaupelJamesW.1979TIoH}. Such heterogeneity is often called frailty \cite{VaupelJamesW.1979TIoH, duchateau_frailty_2008} in the context of survival analysis. Frailty models extend the classic survival models by incorporating random effects (frailties) acting multiplicatively on the baseline hazard function \cite{vaupel_impact_1979, CollettD.1952-author2015Msdi}. In cases where the frailty is greater than one, subjects experience an increased failure hazard. A shared frailty model is a frailty model where the frailties are common or shared among individuals within a cluster or group \cite{CollettD.1952-author2015Msdi, clayton_model_2022, duchateau_frailty_2008, KaragrigoriouAlex2011FMiS, HanagalDD2015Msdu}.

\textcolor{red}{Various parameter estimation methods have been developed for fitting shared frailty models. These methods include penalized partial likelihood (PPL) \cite{TherneauTerryM2003PSMa, duchateau_penalized_2004, ripatti_estimation_2000-1}, expectation-maximization (EM) \cite{DempsterA.P.1977MLfI, KLEINJP1992SEoR}, pseudo full likelihood (PFL) \cite{zucker_pseudo-full_2008-1, GorfineMalka2006Psaw}, hierarchical likelihood (HL) \cite{ha_hierarchical_2022}, maximum marginal likelihood (MML) \cite{vandenBergGerardJ.2016IfSS, lam_marginal_2021-1}, and maximization penalized likelihood (MPL) \cite{JolyP1998APLA, rondeau_maximum_nodate-1} algorithms. These estimation methods have been implemented in various R packages, providing researchers with various options for fitting shared frailty models. The most widely used package for fitting shared frailty models is the \texttt{survival} package \cite{survival-package}, which estimates the parameters by maximizing the penalized partial likelihood. The \texttt{frailtyEM} package \cite{balan_frailtyem_2019-1} implements the general expectation-maximization (EM) algorithm, considering the frailty term as a latent variable. The \texttt{frailtySurv} package \cite{monaco_general_2018-1} adopts a pseudo full likelihood approach for parameter estimation, and the \texttt{frailtyHL} package \cite{ha_frailtyhl_2012-1} estimates the parameters using a hierarchical-likelihood approach. The \texttt{survival}, \texttt{frailtyEM}, \texttt{frailtySurv}, \textcolor{red}{and} \texttt{frailtyHL} packages commonly employed to implement semi-parametric survival models with frailties. For fitting parametric shared frailty models, the \texttt{parfm} package \cite{munda_parfm_2012-1} can be used.  It supports various distributions such as exponential, Weibull, inverse Weibull, Gompertz, lognormal, log-skewNormal, loglogistic, and others. Parameter estimation in \texttt{parfm} uses the maximum marginal likelihood (MML) approach. In addition, the \texttt{frailtypack} package \cite{rondeau_frailtypack_2012-1} fits flexible parametric frailty models. It accommodates scenarios with shared frailty, nested frailty, joint frailty and additive frailty. The parameter estimation in \texttt{frailtypack} is based on the maximization of the penalized log-likelihood.}  


Despite the wide range of parameter estimation methods available in R packages for fitting shared frailty models, it remains unclear if these methods have similar or different performances in terms of precision and efficiency of parameter estimates, computational speed and convergence rate. Early research \cite{HirschKatharina2011Sfss} compared three parameter estimation methods for fitting  shared frailty models through simulation studies. However, with the development of new estimation methods in recent years, an updated comparison is warranted. This study aims to fill this gap by providing a general overview of parameter estimation methods for fitting shared frailty models and comparing their  performances  through simulation studies. Our simulation studies demonstrated that all the parameter estimation methods implemented in the considered R package for fitting shared frailty models yielded very similar and unbiased parameter estimates for the fixed-effect regression coefficients, regardless of sample size, cluster sizes and censoring rates. However, differences were observed in estimating the variance parameter for the frailty term, convergence rate, and computational time. Furthermore, inference for the variance of the frailty terms is not straightforward. Not all  R packages provide an estimation of standard errors for the variance of the frailty terms. Most packages assume the distribution of the estimated variance of the frailty terms is approximately normally distributed. However, since the variance of the frailty terms is positively skewed, a symmetric confidence interval is not ideal. To address this issue, we developed a confidence interval for the variance of the frailty terms and demonstrated its superior performance compared to the conventional confidence intervals provided in some R packages. \textcolor{red}{This improved confidence interval accounts for the skewness of the frailty term distribution and provides a more reliable inference.}
 
The remaining sections of the article are structured as follows. Section \ref{sec2} briefly reviews shared frailty models. Section \ref{sec3} introduces the parameter estimation methods and the corresponding R packages. Section \ref{sec4} presents the design and results of the simulation study for comparing the performance of the parameter estimation methods in the R packages. Finally, the paper concludes with a discussion of the advantages and limitations of each parameter estimation method in the R packages for fitting a shared frailty model in Section \ref{sec5}. Additionally, recommendations for selecting a parameter estimation method for fitting shared frailty models are provided in Section \ref{sec5}.

\section{Shared Frailty Models}\label{sec2}

A shared frailty model is a frailty model where the frailties are common or shared among individuals within groups  \cite{GovindarajuluUshaS.2011FmAt, CollettD.1952-author2015Msdi, HanagalDD2015Msdu}. The formulation of a frailty model for clustered failure survival data is defined as follows. Suppose there are $g$ groups of individuals with $n_i$ individuals in the $i$th group, $i$ = 1, 2, \ldots , $g$. If the number of subjects $n_i$ is 1 for all groups, then the univariate frailty model is obtained \cite{GovindarajuluUshaS.2011FmAt, KaragrigoriouAlex2011FMiS}. Otherwise, the model is called the shared frailty model \cite{CollettD.1952-author2015Msdi, HanagalDD2015Msdu, henderson_analysis_2001, duchateau_frailty_2008, hougaard_frailty_1995} because all subjects in the same cluster share the same frailty value $z_i$. Suppose $t_{ij}$ is the true failure time for the $j$th individual of the $i$th group, which we assume to be a continuous random variable in this article, where $j$ = 1, 2, . . . , $n_i$. Let $t_{ij}^{*}$ denote the realization of $t_{ij}$. In many practical problems, we may not be able to observe $t_{ij}^{*}$ exactly, but we can observe that $t_{ij}$ is greater than a value $c_{ij}$, where $c_{ij}$ be the corresponding censoring time. The observed failure times are denoted by the pair $(y_{ij}, \delta_{ij})$, where $y_{ij}=\min(t_{ij}, c_{ij}), \delta_{ij}=I (t_{ij} < c_{ij})$. The observed data can be written as $y=(y_{11},\cdots,y_{gn_g})$ and $\delta=(\delta_{11},\cdots,\delta_{gn_g}$). This is called right-censoring. \textcolor{red}{Since we only consider right-censoring in this article,} we will use  ``censoring'' as a short for ``right-censoring''.

For the shared frailty models \cite{CollettD.1952-author2015Msdi, HanagalDD2015Msdu}, the hazard of an event at time $t$ for the $j$th individual, $j$ = 1, 2, . . . , $n_i$, in the $i$th group, is then
 \begin{equation}
h_{ij}(t) =z_i \exp(\beta^{T} x_{ij})h_0(t);
\end{equation}
and the survival function for the $j$th individual of the $i$th group at time $t$ follows:
 \begin{equation}
S_{ij}(t) = \exp \bigg\{ -  \int_{0}^{t} h_{ij}(t) \, \mathrm{d}t \bigg \}
=\exp \bigg\{-z_i \exp(\beta^{T} x_{ij}) H_0(t) \bigg \},
\end{equation}
\textcolor{red}{where $x_{ij}$ is a vector of values of $p$ explanatory variables for the $j$th individual in the $i$th group, $\beta$ is the vector of regression coefficients;} $h_0(t)$ is the baseline hazard function, $H_0(t)$ is the baseline  cumulative hazard function, and $z_i$ is the frailty term that is common for all $n_i$ individuals within the $i$th group, let $z=(z_1, \cdots, z_g)$. The hazard and survival functions with frailty effect can also be written as:
\begin{equation}
h_{ij}(t) = \exp(\beta^{T} x_{ij} + u_i)h_0(t),
\end{equation}
and
\begin{equation}
S_{ij}(t) = \exp \bigg\{ - \exp(\beta^{T} x_{ij} + u_i) H_0(t) \bigg \},
\end{equation}
where $u_i$= $\log (z_i)$ is a random effect in the linear component of the proportional hazards model. Note that $z_i$ cannot be negative, but $u_i$ can be any value. \textcolor{red}{If all $u_i$ are equal to zero, the corresponding to $z_i$ is one}, which means the model does not have frailty.  The form of the baseline hazard function is assumed to be unspecified as a semi-parametric model or fully specified to follow a parametric distribution.

In our study, we primarily focus on the shared gamma frailty model \cite{CollettD.1952-author2015Msdi}, as the gamma distribution is commonly used for modelling the frailty effect. The gamma distribution is easy to obtain a closed-form representation of the observable survival, cumulative density, and hazard functions due to the simplicity of the Laplace transform. The gamma distribution is a two-parameter distribution with a shape parameter $k$ and scale parameter $\theta$. \textcolor{red}{The shape parameter determines the shape of the distribution, while the scale parameter influences the spread-out of the distribution. As the value of $k$ varies, the gamma distribution exhibits various shapes.} When $k$ = 1, it is identical to the well-known exponential distribution; when $k$ is large, it takes a bell-shaped form reminiscent of a normal distribution; when $k$ is less than one, it takes exponentially shaped and asymptotic to both the vertical and horizontal axes. Under the assumption $k= \frac{1}{\theta}$, the two-parameter gamma distribution turns into a one-parameter distribution. The expected value is one and the variance is equal to $\theta$. 

\section{Estimation and Inference for Shared Frailty Models}\label{sec3}

\textcolor{red}{In this section, we will provide a brief review of six distinct estimation methods utilized for fitting shared frailty models. These estimation methods vary significantly in terms of the employed likelihoods, the methods for estimating the baseline hazard functions, and the methods for handling the frailty term. }


\subsection{Penalized Partial Likelihood (PPL) Algorithm (R package: \texttt{survival})}\label{sec3.1}

The penalized partial likelihood (PPL) \cite{TherneauTerryM2003PSMa, duchateau_frailty_2008, duchateau_penalized_2004, ripatti_estimation_2000-1, mcgilchrist_reml_1993} approach can be used to estimate parameters in a shared frailty model. This estimation is based on maximizing the penalized partial log-likelihood, which consists of two parts. The first part is the conditional likelihood of the data given the frailties. The second part corresponds to the frailties distribution in which the likelihood is considered a penalty term.  The PPL for the shared frailty model \cite{TherneauTerryM2003PSMa, duchateau_frailty_2008} is then given by
\begin{equation}\label{ppl} 
l_{ppl}(\beta, u, \theta; y, \delta)= l_{part}(\beta, u; y,\delta) + l_{pen}(\theta; u),
\end{equation}
over both $\beta$ and $u$. Here \textcolor{red}{$l_{part}(\beta, u; y,\delta)$} is the partial log-likelihood for the Cox model that includes the random effects. 
\begin{equation}
l_{part}(\beta, u; y,\delta)=  \displaystyle\sum_{i=1}^{g} \displaystyle\sum_{j=1}^{n_{i}} \delta_{ij} \bigg \{ \eta_{ij} - \log \bigg [ \displaystyle\sum_{(q, w) \in R(y_{ij})} \exp(\eta_{qw}) \bigg ] \bigg \},
\end{equation}
 where $\eta_{ij} = \beta^{T} x_{ij}  + u_i $ and $\eta = (\eta_{11},\dots, \eta_{gn_{g}})$.  \textcolor{red}{In the penalty function $l_{pen}(\theta;u)$, the random effect $u_{i}$ is equal to $\log(z_{i})$, where $z_{i}$ usually follow either a lognormal or a gamma distribution.} The penalty function can be written as, 
\begin{equation}
l_{pen}(\theta;u)=  \displaystyle\sum_{i=1}^{g} \log f_{U}({u_{i}\mid\theta}), 
\end{equation}
where $f_{U}({u_{i}})$ denotes the density function of the random effect $u_i$.

The maximization of the PPL consists of an inner and an outer loop\cite{duchateau_frailty_2008, TherneauTerryM2003PSMa}. For the log-normal frailty effects with mean zero and variance $\theta$, the penalized likelihood can be maximized with the Newton-Raphson algorithm in the inner loop. The maximization process proceeds iteratively by starting with a provisional $\theta$ and finding the estimates of the $\beta$'s and the $u$'s that maximize $l_{ppl}(\beta, u, \theta)$. In the outer loop, the restricted maximum likelihood estimator for the $\theta$ is obtained using the best linear unbiased predictors. The process is iterated until convergence. \textcolor{red}{For the gamma frailty effects with unit mean and variance $\theta$, the inner loop is the same as for the log-normal frailty. The outer loop is based on the maximization of a profiled version of the marginal likelihood for $\theta$. Given a specific value of $\theta$, the estimates for $\beta$ and $u$ are determined as the values that maximize the likelihood function $l_{ppl}(\beta, u, \theta)$ with respect to $\theta$. Using these parameter estimates, we can calculate estimates for the baseline hazard function.}

The partial likelihood is not a true likelihood in general, so the maximize the penalized partial likelihood is independent from the baseline hazard function. The Breslow approximation is the first option to estimate the baseline hazard function in nearly all the R packages for fitting Cox regression models with or without frailties. The Breslow estimator \cite{lin_breslow_2007} is the nonparametric maximum likelihood estimation for the cumulative baseline hazard function. It has been implemented in all major statistical software packages. The baseline cumulative hazard function is $H_0(t) = \int_{0}^{t} h_0 (s) \,ds$. Breslow (1972) suggested estimating the cumulative baseline hazard via maximizing likelihood function. After getting the estimators $\hat{\beta}^{T}$ and $\hat{u_i} $, it can provide the nonparametric maximum likelihood estimator of $\hat{H}_0(t)$:
\begin{equation}\label{Breslow}
\hat{H}_0(t)= \displaystyle\sum_{ \{v: y_{(v)} \le t \} } \bigg  \{ \frac{d_{(v)}} { \displaystyle\sum_{(i, j) \in R(y_{(v)}) } \exp(\hat{\beta}^{T}x_{ij}  + \textcolor{red}{\hat{u}_i} )  }  \bigg \},
\end{equation}
where $y_{(1)} < \cdots <  y_{(r)}$ are the ordered distinct event time among the ${y_{ij}}$'s and $R(y_{(v)}) = \{ (i,j): y_{ij} \ge y_{(v)} \}$ is the risk set at $y_{(v)}$ \textcolor{red}{and $d_{(v)}$} is the number of events at $y_{(v)}$.

Arguably the most popular R package for fitting semiparametric shared frailty models is the \texttt{survival} package \cite{survival-package}. The function \texttt{coxph} in the \texttt{survival} package offers a way of fitting shared frailty models via the PPL method. The arguments are the terms including fixed effects of the model,  random effects and the data. The frailty distribution can be  gamma, Gaussian, or t distribution. It accommodates the clustered failures and recurrent events data with right, left, and interval censoring types. When \texttt{coxph} function fits  shared frailty models with clustered failures data,  cluster size should be above five. Otherwise, the random effects will be treated as fixed effects \cite{survival-package}.



\subsection{Expectation-maximization (EM) Algorithm (R package: \texttt{frailtyEM}) }\label{sec3.2}

The expectation-maximization (EM) algorithm \cite{DempsterA.P.1977MLfI, DuchateauLuc2008Tfm} is an iterative method for performing maximum likelihood estimation when the model involves latent variables (missing values). The expectation (E) step attempts to estimate the latent variables via the expectation of the log-likelihood evaluated based on the observed data. The maximization (M) step attempts to optimize the parameters of the model, which computes parameters  by maximizing the expected log-likelihood found in the E step. If we consider the frailty effect $z$ as missing data in the frailty model, the problem can be approached by using EM Algorithm. In the expectation step, compute the unobserved frailties as  the expected values conditional on the observed information and the current parameter estimates are obtained. In the maximization step, we treat these expected values as true information, and new estimates of the parameters of interest are obtained by maximization of the likelihood, given the expected values. 


We first consider the complete data log-likelihood \cite{DempsterA.P.1977MLfI, DuchateauLuc2008Tfm} in which the frailties $z_i$ are regarded as another set of parameters:
\begin{equation}\label{emeq1}
l_{full}(\theta,\beta,z)= \log f(y,\delta,z \mid h_0, \beta,\theta ) = l_{full,1}(\beta;y,\delta,\hat{h}_0) + l_{full,2}(\theta;z), 
\end{equation}
where,
\begin{equation}\label{emeq2}
l_{full,1}(\beta;y,\delta,\hat{h}_0)=  \displaystyle\sum_{i=1}^{g} \displaystyle\sum_{j=1}^{n_{i}}  \bigg  \{ \delta_{ij} \log  \bigg [\hat{h}_0(y_{ij}) z_i \exp(\beta^{T} x_{ij})  \bigg ] - \hat{H}_0(y_{ij}) z_i \exp(\beta^{T} x_{ij}) \bigg  \}
\end{equation}
and
\begin{equation}\label{emeq3}
 l_{full,2}(\theta; z) =  \displaystyle\sum_{i=1}^{g} \log f_{Z}({z_{i}}\mid\theta).
\end{equation}
We use $l_{full,1}(\beta;y,\delta,\hat{h}_0)$ to estimate $\beta$, and $ l_{full,2}(\theta; z)$ to estimate $\theta$. Within the framework of the EM algorithm, the expected value of the full log-likelihood needs to be maximised \cite{DempsterA.P.1977MLfI, DuchateauLuc2008Tfm}. In the E step, the “posterior” distribution of the frailties $p(z_i \mid y_i, \delta_i, \beta^{(k-1)}, \theta^{(k-1)})$ can be obtained. Then, the $E^{(k)}(z_i)$ and $E^{(k)}(\log z_i)$ can be calculated. In the M step, the loglikelihood in \eqref{emeq1} is profiled to a partial loglikelihood by considering the frailties as fixed offset terms, then the $E^{(k)}(z_i)$ and $E^{(k)}(\log z_i)$ are considered to be the true value to replace the $z_i$'s and $\log z_i$'s in the partial loglikelihood leading to 

\begin{equation}\label{emeq4}
l_{part,1}^{(k)}(\beta)=  \displaystyle\sum_{i=1}^{g} \displaystyle\sum_{j=1}^{n_{i}} \delta_{ij}  \bigg  \{ E^{(k)}(\log z_i) + \beta^{T} x_{ij} - \log  \bigg ( \displaystyle\sum_{(q, w) \in R(y_{ij})} E^{(k)}(z_q) \exp(x_{qw} \beta )  \bigg )  \bigg   \}.
\end{equation}
The new estimates $\beta^{(k)}$ can be obtained from the $l_{part,1}(\beta)$. A new estimate $\theta^{(k)}$ can be obtained immediately by maximization of $ l_{full,2}(\theta; u)$, replacing $z_i$'s and $\log z_i$'s in \eqref{emeq3} by the current expected values at iteration step $k$. \textcolor{red}{ The Breslow estimator in equation \eqref{Breslow} is applied to estimate the baseline hazard function}, which is required in the expectation step. In the initialization E step, $\theta^{(0)}$ is set to one and an ordinary Cox model is fitted leading to estimates $\beta^{(0)}$. Next, we iterate between the expectation and maximization steps until convergence. The marginal loglikelihood can be used for assessing the convergence of the algorithm

The \texttt{frailtyEM} package was written by Theodor et al. \cite{balan_frailtyem_2019-1}. It provides maximum likelihood estimation of semiparametric shared frailty models using the expectation-maximization algorithm. The main model fitting function in frailtyEM is emfrail,  and the user has to define the main arguments formula, data set, distribution and control. This formulation is common to most survival analysis packages, allowing for several scenarios, including possibly left truncated clustered failures and recurrent events in both calendar time and gap time formulation. The distribution argument determines the frailty distribution; the gamma, stable and power variance function family distributions are supported. The control argument can be provided by the emfrail's control() function, and it controls parameters for the emfrail. The package can access predicted survival and cumulative hazard curves, both for an individual and on a population level. The results from \texttt{frailtyEM} package are very close to the survival package.

\subsection{Maximum Marginal Likelihood (MML) Algorithm (R package: \texttt{parfm}) }\label{sec3.3}

The maximum marginal likelihood (MML) approach was proposed for estimating the parameters for  shared frailty models \cite{vandenBergGerardJ.2016IfSS, lam_marginal_2021-1}. The frailties are integrated out by averaging the conditional likelihood with respect to the frailty distribution. This method can be applied to any frailty distribution with explicit Laplace transform. 

For the right-censored clustered survival data, the observation for the $j$th individual in the $i$th group are the triple $(y_{ij}, \delta_{ij}, x_{ij})$. Further, if left-truncation is also present, truncation times $\tau_{ij}$ are gathered in the vector $\tau$, i.e., $\tau=(\tau_{11}\cdots,\tau_{gn_g})$. Let $\psi$ represent a vector of parameters for the baseline hazard function. The marginal log-likelihood \cite{vandenBergGerardJ.2016IfSS, lam_marginal_2021-1} can be written as 
\begin{equation} \label{mmleq1}
\begin{aligned}
l_{marg}(\psi, \beta,\theta; y,\delta,\tau,x) = \displaystyle\sum_{i=1}^{g}  \bigg \{  \bigg [  \displaystyle\sum_{j=1}^{n_{i}}  \delta_{ij} (\log(h_0(y_{ij}\mid \psi))+ \beta^{T} x_{ij}) \bigg ] \\
+ \log \bigg [(-1)^{d_i} \mathcal{L}^{d_i} ( \displaystyle\sum_{j=1}^{n_{i}} H_0(y_{ij}\mid\psi)  \exp(\beta^{T} x_{ij}) ) \bigg ] \\
- \log \bigg [ \mathcal{L}( \displaystyle\sum_{j=1}^{n_{i}} H_0(\tau_{ij}\mid \psi) \exp(\beta^{T} x_{ij}) ) \bigg] \bigg \},
\end{aligned}
\end{equation}
where $\theta$ is used as the vector of parameters for the frailty distribution function, $d_i = \displaystyle\sum_{j=1}^{n_{i}}  \delta_{ij}$ the number of events in the $i$-th cluster, and $ \mathcal{L}^{q} (\cdot)$ is the q-th derivative of the Laplace transform of the frailty distribution, which is defined as,
\begin{equation}
\mathcal{L}(s)=  \int_{0}^{\infty} \exp (-sz ) f(z) \,dz,
\end{equation}
where $f(z)$ is the density function of frailty term $z$. If the higher-order derivatives $ \mathcal{L}^{q} (\cdot)$ of the Laplace transform up to $q= \max \{d_1, \dots, d_G \}$ are able to compute, the estimates of $\psi$, $\beta$, $\theta$, can be obtained by maximising the marginal log-likelihood \eqref{mmleq1}. \textcolor{red}{The parametric estimation approach is applicable for modelling the form of the baseline hazard.}

The \texttt{parfm} package \cite{munda_parfm_2012-1} estimates the parameters for  parametric frailty models by maximizing the marginal log-likelihood. The  baseline hazard distributions can be exponential, Weibull, inverse Weibull (Frechet), Gompertz, lognormal, log-kewNormal, and loglogistic. The frailty distribution can be  gamma, positive stable, inverse Gaussian, and lognormal distribution.

\subsection{Hierarchical Likelihood (HL) Algorithm (R package: \texttt{frailtyHL}) }

Lee \& Nelder \cite{LeeY.1996HGLM} proposed the use of hierarchical likelihood for fitting the model with random effects. The hierarchical likelihood consists of data, parameters and unobserved random effects. This method can avoid the integration over the random-effect distributions.  The method  is the statistically efficient estimation in frailty models by using the Laplace approximation. Thus, the h-likelihood can be used directly for inference on random effects. 

For the observe  $y_{ij}$ and the censoring indicator is $\delta_{ij}$, the h-likelihood \cite{LeeY.1996HGLM} for a frailty model is defined by
\begin{equation} \label{hleq1}
hl(\beta, \theta, u; y, \delta, \hat{h}_0) = l_0( \beta; y,\delta,z, \hat{h}_0)  + l_1(\theta; u),
\end{equation}
where $l_0$ is the sum of conditional log densities for ($y, \delta$) given the random effect $u= (\log z_1, \cdots, \log z_g)$; then it follows:
\begin{equation}\label{hleq2}
\begin{aligned}
l_0( \beta; y,\delta, u, \hat{h}_0) &= \displaystyle\sum_{ij} \log f(y_{ij}, \delta_{ij} \mid \beta,u_{i} , \hat{h}_0) \\
&= \displaystyle\sum_{ij} \delta_{ij}  \bigg \{ \log \hat{h}_0(y_{ij}) +(\beta^{T} x_{ij} + u_i) \bigg\} \\
&- \displaystyle\sum_{ij} \bigg\{ \hat{H}_0(y_{ij}) \exp(\beta^{T} x_{ij}+ u_i) \bigg\} \\
\end{aligned}
\end{equation}
$l_1$ is the sum of log densities for random effects $u$ with parameter $\theta$, which is defined by 
\begin{equation}\label{hleq3}
l_1(\theta; u) = \displaystyle\sum_{i} \log f_U(u_i\mid \theta).
\end{equation}

\textcolor{red}{The Breslow estimator in equation \eqref{Breslow} is employed to estimate the baseline hazard function $\hat{h}_0$.} From the equation \eqref{ppl}, the penalized partial likelihood is defined as,
\begin{equation} \label{hleq4}
\begin{aligned}
l_{ppl}(\beta, u, \theta; y, \delta) &=  \displaystyle\sum_{ij} \delta_{ij} \bigg \{ (\beta^{T} x_{ij} + u_i) - \log \bigg [ \displaystyle\sum_{(q, w) \in R(y_{ij})} \exp(x_{qw} \beta + u_q) \bigg ] \bigg \} \\
&+ \displaystyle\sum_{i} \log f_{U}({u_{i}\mid \theta}).
\end{aligned}
\end{equation}
The papers \cite{HaIlDo2001Hlaf,HAILDO2010BRoL} showed  that $hl(\theta,\beta)$ is equal to the $l_{ppl}(\beta, u, \theta; y, \delta)$ plus a constant,
\begin{equation} \label{hleq5}
hl(\beta,u,\theta) =l_{ppl}(\beta, u, \theta; y, \delta) + \displaystyle\sum_{ (q,w) \in R(y_{ij}) } d_{qw} \bigg \{ \log \hat{h}_0(y_{qw}) -1 \bigg \},
\end{equation}
where $\displaystyle\sum_{ (q,w) \in R(y_{ij}) } d_{qw} \bigg \{ \log \hat{h}_0(y_{qw}) -1 \bigg \}$ is a constant and $d_{qw}$ is the number of element in the risk set $R(y_{qw})$. \textcolor{red}{Accordingly, given the frailty parameter $\theta$, the hierarchical likelihood methods for estimating the parameter estimator $\beta$ can be obtained by maximizing the profile marginal likelihood after eliminating $H_0(t)$.} The Laplace approximation can be used when the marginal likelihood is hard to obtain. Given $\hat{\beta}$ and $\hat{u}$, the maximum adjusted profile hierarchical likelihood for estimating the variance of the frailty terms $\theta$ can be obtained. We iterate these steps until convergence. The estimates of the standard errors can be computed \cite{LeeY.1996HGLM}. 

The \texttt{frailtyHL} package created by Ha et al. \cite{ha_frailtyhl_2012-1} implements the hierarchical-likelihood procedures for fitting semi-parametric frailty models with non-parametric baseline hazards. The package fits shared or multilevel frailty models for correlated survival data. The lognormal or gamma distributions can be adopted as the frailty distribution, corresponding to the normal or log-gamma distributions for the log frailties. The results of estimates of fixed effects, random effects, and variance components as well as their standard errors are provided. In addition, it provides a statistical test for the variance components of frailties and also three AIC criteria for the model selection. However, the package does not provide the interval estimation of frailty. 

\subsection{Pseudo Full Likelihood (PFL) Algorithm (R package: \texttt{frailtySurv})}

Pseudo full likelihood \cite{zucker_pseudo-full_2008-1, GorfineMalka2006Psaw} is a new method that can handle any parametric frailty distribution with finite moments. A simple univariate numerical integration can deal with non-conjugate frailty distributions. The cumulative hazard function is estimated via a noniterative procedure. Other properties follow the consistency and asymptotic normality of the parameter estimates and a direct, consistent covariance estimator. It is easy to compute and implement. From the study of Gorfine et al. \citep{GorfineMalka2006Psaw}, the results of estimation for fitting the shared frailty model are very similar to the EM-based method.

In the shared frailty model, we assume further that the observed data consisting of $y, \delta, x$ are independent. The proposed approach can estimate the regression coefficient vector $\beta$, the frailty distribution’s parameter $\theta$, and the non-parametric cumulative baseline hazard $H_0$. Let $\tau$ be the end of the observation period. The full likelihood \cite{zucker_pseudo-full_2008-1, GorfineMalka2006Psaw} can be defined as
\begin{equation} \label{ }
L(\beta,\theta, H_0) =  \prod_{i=1}^{g} \prod_{j=1}^{n_i}  \bigg \{ h_0( y_{ij}) \exp(\beta^{T} x_{ij}) \bigg \} ^{\delta_{ij}}  \prod_{i=1}^{g} (-1)^{N_{i.}(\tau)} \mathcal{L}^{(N_{i.})} \{ H_{i.}(\tau) \},
\end{equation}
where $N_{ij}(t)= \delta_{ij} I(y_{ij} \le t)$, $N_{i.}(t)=\sum_{j=1}^{n_i} N_{ij}(t) $, $H_{ij}(t) = H_0(\min \{y_{ij},t \}) \exp(\beta^{T} x_{ij})$, $H_{i.}(t)=\sum_{j=1}^{n_i} H_{ij}(t) $, $\mathcal{L}$ is the Laplace transform of the frailty distribution and  $\mathcal{L}^{(m)}, m=1, 2,  \cdots$ are the $m$th derivatives of $\mathcal{L}$. Note that the $m$th derivatives of the Laplace transform evaluated at $H_{i.}(\tau)$ equals to $(-1)^{N_{i.}(\tau)} \int z^{N_{i.}(\tau)} \exp \{ -z H_{i.}(\tau) \} f(z) \,dz$. The log-likelihood equals to

\begin{equation} \label{}
l(h_0, \theta,\beta) = \displaystyle\sum_{i=1}^{g} \displaystyle\sum_{j=1}^{n_{i}} \bigg \{ \delta_{ij}  \log \{ h_0(y_{ij}) \exp(\beta^{T} x_{ij}) \bigg \}+\displaystyle\sum_{i=1}^{g} \log \mathcal{L}^{(N_{i.})} \{ H_{i.}(\tau) \}.
\end{equation}
Obviously, an estimator of $H_0$ is required in the log-likelihood function to obtain estimators of $\beta$ and $\theta$. In the initialization step, $\theta$ should be set as a value and a standard Cox model is fitted to obtain initial estimates of $\beta$. For given these two initial values, $H_0$ is estimated via the Breslow estimator with jumps at the ordered observed failure times $\tau_v$, $v=1,\cdots,r$. The detailed step of the  baseline hazard estimation is referred to by Gorfine et al. \cite{GorfineMalka2006Psaw}. Then, $\hat{H}_0$ is substituted into the log-likelihood function. The estimators of $\hat{\beta}$ and $\hat{\theta}$ can be obtained by maximizing the log-likelihood function. Iterate these steps until convergence.

The R package \texttt{frailtySurv} \cite{monaco_general_2018-1} can be used for simulating and fitting semi-parametric shared frailty models. It can be applied for various frailty distributions, including gamma, log-normal, inverse Gaussian and power variance functions via pseudo full likelihood approach. The parameters’ estimators are consistent and asymptotically normally distributed. The results of this package can be performed using the normal distribution, such as hypothesis testing and confidence intervals. Only right-censoring with clustered failures dataset is supported by \texttt{frailtySurv}.

\subsection{Maximization Penalized Likelihood (MPL) Algorithm (R package: \texttt{frailtypack}) }

The maximum penalized likelihood estimation \cite{JolyP1998APLA, rondeau_maximum_nodate-1} can be applied to the nonparametric estimation of a continuous hazard function in a shared frailty model. This approach is based on the penalized full likelihood, which is opposed to the penalized partial likelihood. We assume that the frailty effects are distributed from a gamma distribution with mean 1 and variance $\theta$. For the observe $y$, $\delta$, and the truncation times $\tau$, the full marginal loglikelihood for the shared gamma frailty model has an analytical formulation \cite{KLEINJP1992SEoR}
\begin{equation}
\begin{aligned}
 l(\beta, \theta, h_0)=  \displaystyle\sum_{i=1}^{g} \bigg \{  \bigg [ \displaystyle\sum_{j=1}^{n_{i}}  \delta_{ij} \log h_0(y_{ij}) \bigg ] - (\frac{1}{\theta} +m_i) \log \bigg[ 1+ \theta \displaystyle\sum_{j=1}^{n_{i}}  H_0(y_{ij}) \bigg] \\
 + \frac{1}{\theta} \log \bigg [ 1+ \theta \displaystyle\sum_{j=1}^{n_{i}}  H_0(\tau_{ij}) \bigg ]  + I({m_i \neq 0 } ) \displaystyle\sum_{k=1}^{m_{i}} \log \bigg (1+ \theta(m_i - k) \bigg ) \bigg \}
\end{aligned}
 \end{equation}
where the number of recurrent events is $m_i = \displaystyle\sum_{j=1}^{n_{i}} \delta_{ij} $.

The penalized loglikelihood function for the shared gamma frailty model \cite{JolyP1998APLA, rondeau_maximum_nodate-1}  follows
\begin{equation}
 pl(\beta, \theta, h_0)= l(\beta, \theta, h_0) - k \int_{0}^{\infty} {h_0}^{\prime \prime} (t)^2 \,dt
 \end{equation}
where $k$ is a positive smoothing parameter  that controls the trade-off between the data fit and the smoothness of the functions. The smoothing parameter needs to be a fixed value, and the estimators of $\beta$ and $\theta$ can be obtained via the maximization of the penalized likelihood. The robust Marquardt algorithm \cite{MarquardtDonaldW.1963AAfL} is used to estimate parameters, which is a combination between a Newton Raphson algorithm and the steepest descent algorithm. The estimator of the baseline hazard function $h_0(\cdot)$ can be approximated on the basis of Cubic M-splines with Q knots \cite{RamsayJ.O.1988MRSi, emura_programs_nodate}. The splines, the regression coefficients, and the variance of the frailty term are initialized to 0.1 in the shared frailty model. The model can be fit firstly, then adjusted Cox model to give new initial values for the splines and the regression coefficients.

The \texttt{frailtypack} package \cite{rondeau_frailtypack_2012-1} allows fitting Cox models and four types of frailty models (shared, nested, joint, additive). The function \texttt{frailtyPenal} fits the shared  frailty model by using the MPL method with the splines to estimate the baseline hazard. As a result, due to the use of splines with a specified number of knots for modelling the baseline hazard, this approach can be considered a parametric model.
\textcolor{red}{According to the reference manual, the baseline hazard can be modelled using either a piecewise constant function or Weibull functions. However, it is important to note that the default function for the baseline hazard in the frailtyPenal is splines.} Right-censored or left-truncated data are considered in this package. The arguments are the terms including the fixed effect, the cluster variable, and the data set. In addition, there are two arguments in the formula that need to be specified: n.knots (4 up to 20) and kappa1 (smoothing parameter). \textcolor{red}{In our simulation, we employed the splines function with 15 knots and set the value of kappa1 to 1. Additionally, we utilized the argument cross.validation, a logical value that must be set to `true'.}




\textcolor{red}{Table \ref{method} presents a concise overview of the six parameter estimation methods mentioned above, specifically focusing on their distinctions in terms of likelihood, baseline hazard form, and the approach used to handle frailty terms. This table provides a clear visualization of the similarities and differences among the various methods, allowing for easy comparisons and assessments.}

\begin{table}[htb]
\footnotesize
\addtolength{\tabcolsep}{0pt}
\caption{Six parameter estimation methods for fitting shared frailty models in terms of likelihood, the form of baseline hazard, and the method for handling frailty terms. PPL = penalized partial likelihood, MML = maximum marginal likelihood, EM = expectation maximization, PFL = pseudo full likelihood, HL = h-likelihood, MPL = maximization penalized loglikelihood. }
\begin{tabular}{|l | l | l | l | l | }
 \hline
Package & Algorithm&Likelihood &Form of $H_0$  & Methods for handling $z$   \\
\hline
survival &PPL& Partial likelihood  & Step function  &Penalization \\
\hline
parfm &MML &Parametric full likelihood  &\shortstack[l]{Parametric \\distributions } & Laplace transform\\
\hline
frailtyEM &EM & Partial likelihood &Step function  &EM Algorithm \\
\hline
frailtySurv &PFL&Full likelihood  &Step function  &Laplace transform \\
\hline
frailtyHL &HL& Partial likelihood  &Step function  &Laplace approximation  \\
\hline
frailtypack &MPL&Full likelihood  &Spline  &Integration \\
\hline
\end{tabular}
\label{method}
\end{table}

Table \ref{tab:package} provides a summary of  the above-mentioned six R packages for fitting shared frailty models in terms of the  frailty distribution, algorithm, censoring type and data type. 

\begin{table}[htb]
\footnotesize
\addtolength{\tabcolsep}{0pt}
\caption{R packages for fitting shared frailty models in terms of the primary R function,  frailty distribution, fitting algorithm, censoring type and data type. PPL = penalized partial likelihood, MML = maximum marginal likelihood, EM = expectation maximization, PFL = pseudo full likelihood, HL = h-likelihood, MPL = maximization penalized loglikelihood. }
\begin{tabular}{| l | l | l | l | l | l |}
 \hline
 Package&Function &Frailty distribution  & Algorithm& Censoring & Data \\
\hline
survival&coxph  &\shortstack[l]{Gamma, \\Log-normal, t}  & PPL&\shortstack[l]{Right, \\interval, \\Left} &\shortstack[l]{Clustered failures, \\Recurrent events}\\
\hline
parfm & parfm &\shortstack[l]{Gamma, \\Log-normal,\\ Positive Stable, \\Inverse Gaussian} &MML& Right&\shortstack[l]{Clustered failures,\\ Left truncation}\\
\hline
frailtyEM &emfrail &\shortstack[l]{Gamma, \\Positive Stable, \\ Inverse Gaussian, \\Compound Poisson,\\ Power Variance \\Function }&EM &Right &\shortstack[l]{Clustered failures,\\Recurrent events,\\ Left truncation}\\
\hline
frailtySurv&fitfrail  &\shortstack[l]{Gamma,\\ Log-normal, \\Inverse Gaussian, \\Power Variance \\Function}&PFL&Right &Clustered failures\\
\hline
frailtyHL&frailtyHL  & \shortstack[l]{Gamma, \\Log-normal}  &HL &Right &Clustered failures\\
\hline
frailtypack& frailtyPenal & \shortstack[l]{Gamma,\\ Log-normal}&MPL&Right &\shortstack[l]{Clustered failures,\\ Recurrent events,\\ Left truncation, \\Correlated structure}\\
\hline
\end{tabular}
\label{tab:package}
\end{table}

\section{Simulations and Results } \label{sec4}

We conducted simulation studies to investigate \textcolor{red}{the performances of parameter estimation methods implemented in R packages for fitting the shared frailty models.} We generated the true  failure time  from a Weibull regression model with shape  parameter ($\alpha=3$) and scale parameter  ($\lambda=0.007$) \cite{HirschKatharina2011Sfss}. More specifically $ t_{ij}= \{ - \log(u_{ij})/[\lambda \exp({\beta_1 }x_{ij}^{(1)} +{\beta_2} x_{ij}^{(2)} +{\beta_3} x_{ij}^{(3)} + z_i)]\}^{(1 / \alpha)}$, where $i$ = \{1,$\cdots$, g\}, $j $ = \{1,$\cdots$, $n_i$\} and $u_{ij}$ was simulated from Uniform[0, 1]. The censoring time $C_{i}$ was simulated from an exponential distribution, $\exp(\theta)$, where $\theta$ was set to obtain three different censoring rates ($c$): $20\%, 50\%$, and $80\%$, respectively. Three covariates were generated including $x_{ij}^{(1)}$ from a Uniform[0, 1], $x_{ij}^{(2)}$ from a Normal(0, 1), and  $x_{ij}^{(3)}$ from a Bern(0.25).  We set true regression parameters for the three covariates as ${\beta_1}=1$, ${\beta_2}=-1$, ${\beta_3}=0.5$, respectively. The frailty term was generated from a gamma distribution with a variance of 0.5. \textcolor{red}{All the parameter settings are consistent with the previous study\cite{HirschKatharina2011Sfss}.} We considered fitting a shared frailty gamma model assuming $h_{ij}(t_{ij}) =z_i \exp(\beta_1 x_{ij}^{(1)} + \beta_2 x_{ij}^{(2)} +\beta_3 x_{ij}^{(3)} )h_0(t_{ij})$ as a true model. \textcolor{red}{Via investigating if the performances of parameter estimation methods} depend  on  sample size, we simulated datasets with varying sample sizes $n$ ranging from 100 to 800. For a sample of size 100, the observations were grouped into 10 clusters of size 10. For a sample of size 400, the observations were grouped into 10 clusters of size 40 or 40 clusters of size 10. For a sample of size 800, the observations were grouped into 10 clusters of size 80 or 80 clusters of size 10. All considered parameter estimation methods available in R  packages were applied to the same simulated dataset in each scenario. Using 1000 datasets generated under each scenario, we examined the precision of the parameter estimates in terms of bias and standard errors of the estimated parameters, as well as the coverage probability (CP) of the estimated parameters. We also investigated the performance of the packages in terms of convergence rate and average computing time under each simulation scenario. \textcolor{red}{In our comparative study, we will use the name of the R package to represent each estimation method.}

%
\subsection{Estimated parameters}
\begin{figure}
\includegraphics[width=\textwidth, height=\textwidth]{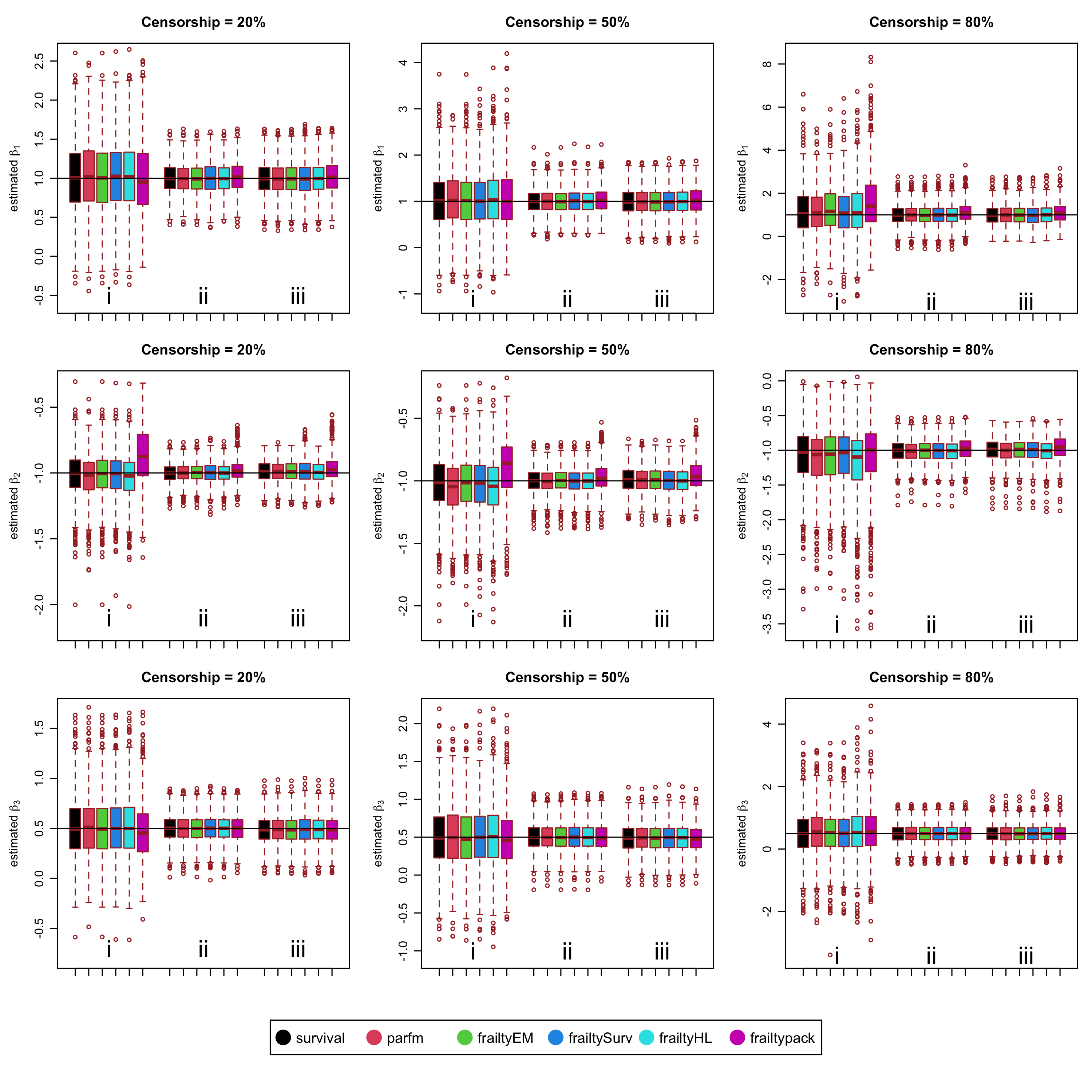}

\caption{The estimated regression coefficients over 1000 samples simulated from the true model. True values of the regression coefficients are indicated as horizontal lines. The  first, second and third columns correspond to 20\%, 50\% and 80\% censoring rates, respectively. In each panel, the left group corresponds to scenario (i) with 10 clusters of size 10; the middle group corresponds to scenario (ii) with 10 clusters of size 40; the right group corresponds to scenario (iii) with 40 clusters of size 10}
\label{fig: fixed_effect}
\end{figure}

Figure \ref{fig: fixed_effect} \textcolor{red}{presents the estimated regression coefficients over 1000 repeated samples when the sample size $n$ was 100 and 400. The results indicate that all packages performed similarly in estimating the regression coefficients. However, it should be noted that  \texttt{frailtypack} slightly overestimated $\beta_2$ for the sample size of 100}. Not surprisingly, as the censoring rate increases, the estimated regression coefficients are subject to more variability. 

\begin{figure}
\includegraphics[width=1\textwidth, height=0.6\textwidth]{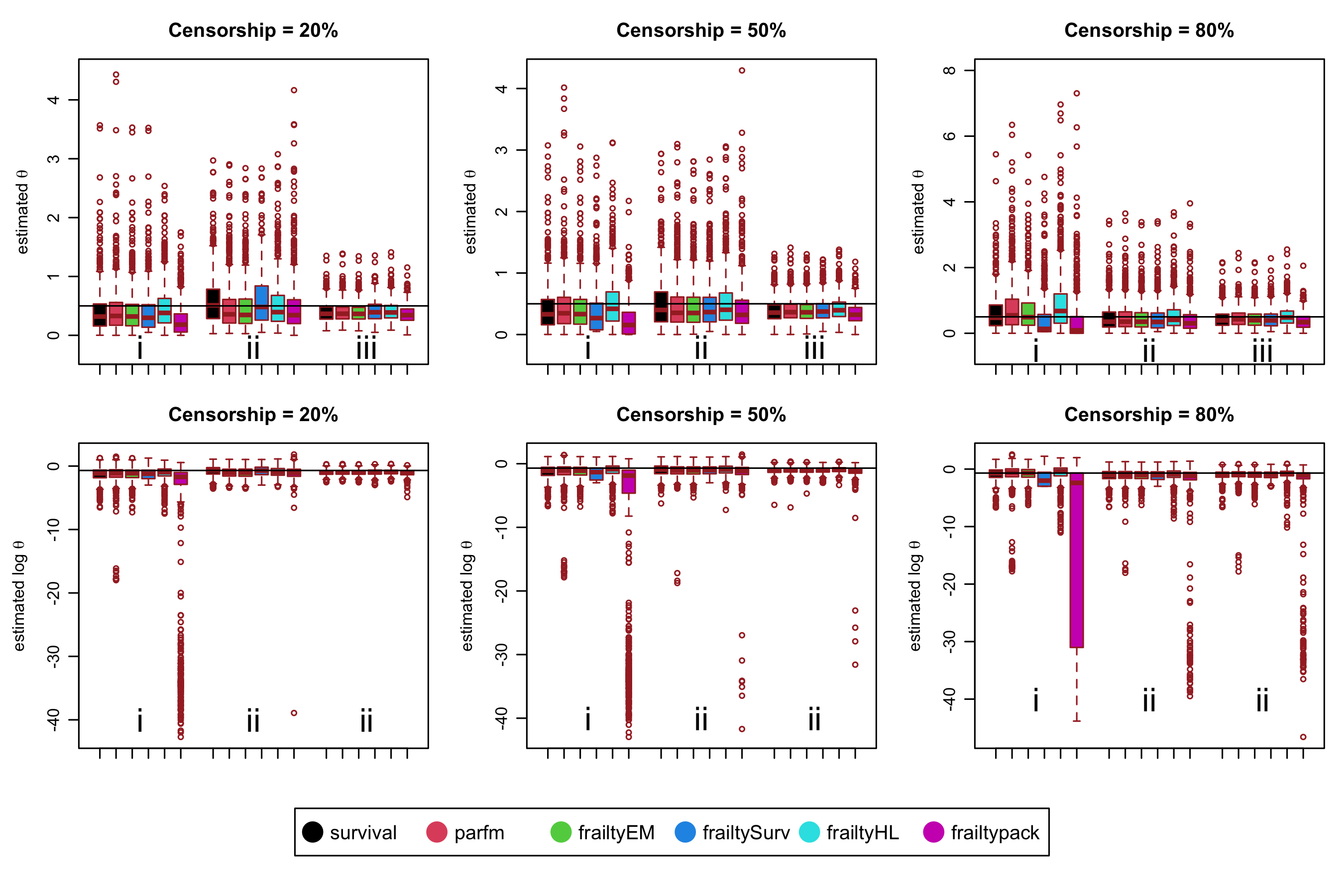}

\caption{Boxplots of the estimated variance of the frailty term over 1000 samples simulated from the true model. True values of the variance parameters are indicated as horizontal lines. The  first, second and third columns correspond to 20\%, 50\% and 80\% censoring rates, respectively. In each panel, the left group corresponds to scenario (i) with 10 clusters of size 10; the middle group corresponds to scenario (ii) with 10 clusters of size 40; the right group corresponds to scenario (iii) with 40 clusters of size 10}
\label{fig: random_effect}
\end{figure}

As displayed in the top panels of Figure \ref{fig: random_effect}, the estimated variance parameter of the frailty term was underestimated. This underestimation was more prominent when there were 10 clusters of size 10 and 10 clusters of size 40, compared to 40 clusters of size 10. These findings suggest that a lower number of clusters results in higher variability in estimating the variance parameter of the frailty term. The distribution of the variance parameter of the random effect term is known to be positively skewed \cite{McCullochCharlesE.2011MtSo}. To enhance the visualization of the estimated variance of the frailty terms, the log-transformed values were presented in the bottom panels of Figure \ref{fig: random_effect}. The Figure demonstrates that the \texttt{frailtypack} package yielded numerous extremely small estimates for the variance parameter of the frailty term. This trend is particularly noticeable for a sample size of 100 and also for a high censoring rate in a sample size of 400.
 
Tables \cref{table_100_20,table_100_50,table_100_80,table1,table2,table3} in the Appendix present detailed information about parameter estimates including the bias, mean and median of the standard error, empirical standard error and mean square error (MSE) of the estimated model parameters when the total sample size is 100 and 400 with the percentage of censoring rate is 20$\%$, 50$\%$, and 80$\%$, respectively. The \texttt{survival} package  does not provide the estimated  standard error of the variance of the frailty terms. The empirical standard errors for the regression coefficients and variance parameter are defined based on their point estimates over simulated samples, which are calculated as Empirical SE=$ \sqrt{\frac{\sum_{i=1}^{n_{sim}} (\hat{\theta_i}-\bar{\theta} )^2}{n_{sim}-1}} $, where $n_{sim}$ is the number of successful fittings to the 1000 datasets, and $\theta$ donates the true regression coefficient or variance parameter of the frailty. The results indicate that when the censoring rate increases, the variance of the frailty terms estimate has a smaller bias but larger variability. This finding is in line with previous research \cite{HirschKatharina2011Sfss, petersen_inference_2006}. The underestimation was even observed in the settings without censoring. \textcolor{red}{The maximum likelihood variance estimator in linear mixed models has a tendency to underestimate the true variance. This discrepancy arises because an unknown mean estimate is used in the calculation of the variance estimates. Similarly, when estimating the variance of the frailty term in the shared frailty model using the maximum likelihood method, it also tends to be underestimated. This similarity suggests that the underestimation of the frailty term variance in the shared frailty model may share a similar reason with the underestimation observed in the linear mixed model. Consequently, the underestimation of the frailty term variances could be one of the reasons why the survival package does not provide a standard error for it.} The mean and median of the standard errors provided by all packages are very close to the empirical standard errors. 

\begin{figure}
\includegraphics[width=\textwidth, height=\textwidth]{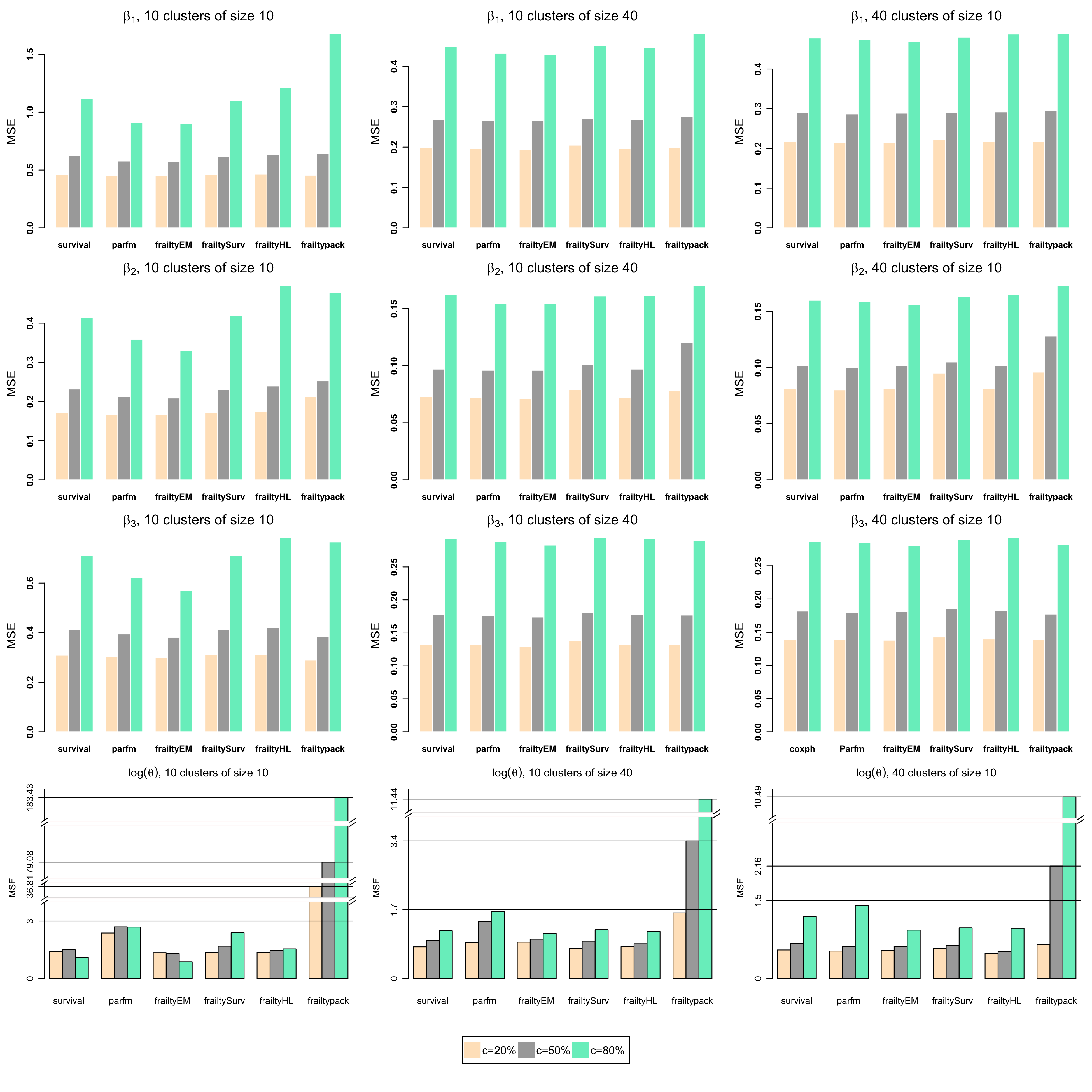}

\caption{Comparison of the MSEs of estimated regression coefficients and the log variance of the frailty terms. The first, second, third and fourth rows correspond to the results for $\beta_1$, $\beta_2$, $\beta_3$ and $\theta$, respectively. The left panels correspond to a scenario with 10 clusters of size 10, the middle panels correspond to a scenario with 10 clusters of size 40 and the right panels correspond to the scenario with 40 clusters of size 10. In each  panel, the yellow, gray and green bars correspond to 20\%, 50\% and 80\% censoring rates, respectively.}
\label{fig:MSE}
\end{figure}

Figure \ref{fig:MSE} displays the MSEs of all the parameters in the scenario of 10 clusters of size 10  (left panels), 10 clusters of size 40 (middle panels) and 40 clusters of size 10 (right panels). The results of MSEs for all the estimated regression coefficients indicate that as the percentage of censoring increases, the MSEs of the estimated regression coefficients increased for all the R packages. However, the \texttt{frailtypack} had slightly larger MSEs for $\beta_2$ compared to other packages. \textcolor{red}{The fourth row of Figure \ref{fig:MSE} shows the results of MSEs for the log-transformed variance of the frailty terms. The MSE of the variance of the frailty terms increases for most R packages as the percentage of censoring increases, but the MSE of 20\% and  50\% censoring are larger than that 80 \% censoring for \texttt{survival} and \texttt{frailtyEM} package when the sample size $n$ was 100. Moreover, the MSEs of the \texttt{frailtypack} package are much larger compared to the other package, especially in the case of sample size 100.}

\subsection{Coverage Probability (CP)}
\begin{figure}[hp]
\includegraphics[width=\textwidth, height=\textwidth]{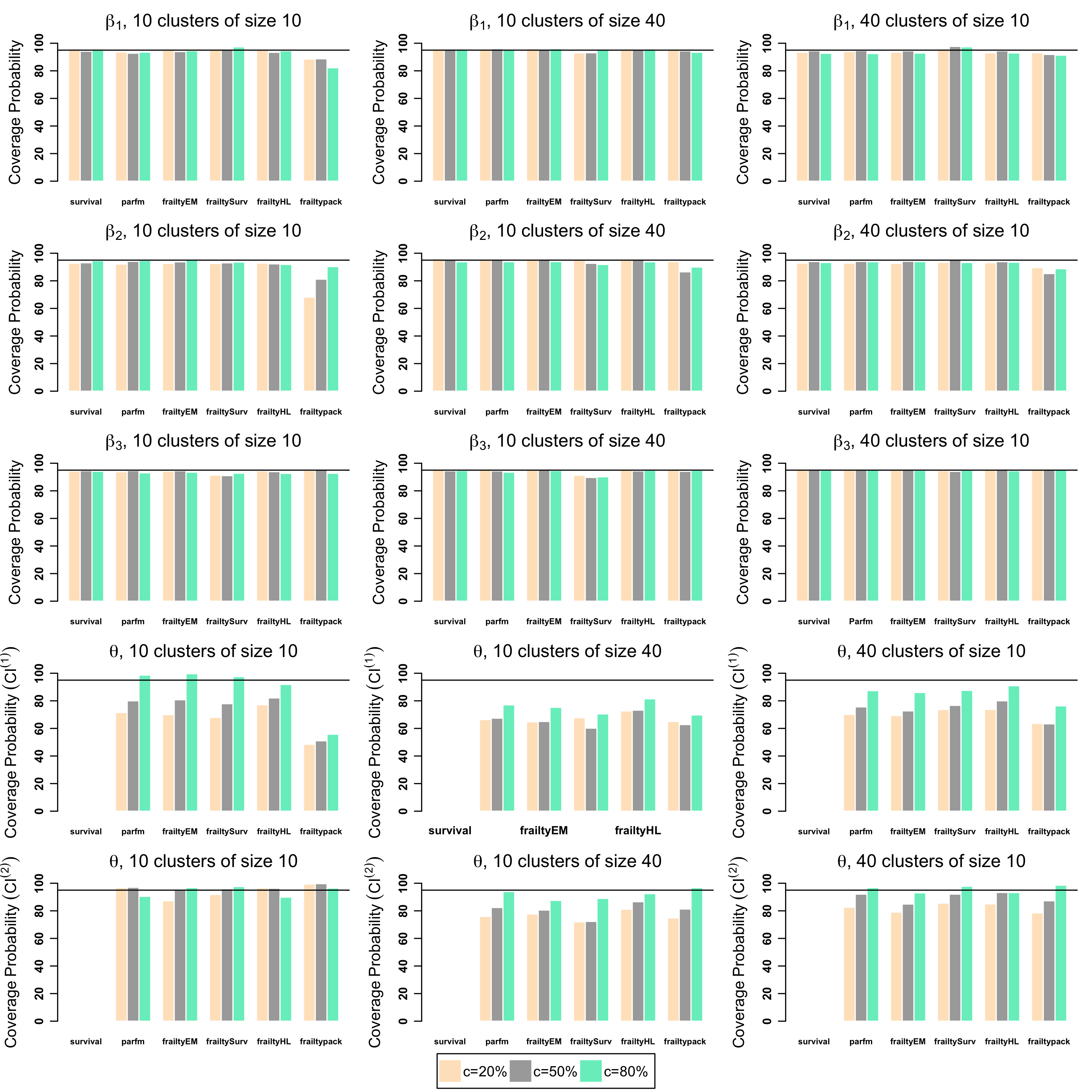}
\caption{The coverage probability of the  95\%  confidence interval  for each  estimated regression coefficient and the variance of the frailty terms. The black horizontal line  indicates the 95\% nominal level. The first, second and third rows correspond to the results for $\beta_1$, $\beta_2$ and $\beta_3$, respectively. The fourth and  fifth rows correspond to CI$^{(1)}$ under normal approximation and CI$^{(2)}$ under log transformation as described in section 4.2. The left panels correspond to a scenario with 10 clusters of size 10, The middle panels correspond to a scenario with 10 clusters of size 40 and the right panels correspond to the scenario with 40 clusters of size 10. In each  panel, the yellow, gray and green bars correspond to 20\%, 50\% and 80\% censoring rates, respectively.}
\label{fig: cp}
\end{figure}

For all the R packages considered in this paper, the 95\% confidence intervals (CI) of the regression coefficients are calculated based on normal approximation, i.e., $\hat{\beta} \pm 1.96* SE(\hat{\beta})$. The first, second, and third rows of Figure \ref{fig: cp} displays the results of the coverage probabilities (CPs) for the 95\% confidence intervals (CIs) of three regression coefficients in different scenarios. The left panels correspond to the scenario of 10 clusters of size 10, the middle panels depict the scenario of 10 clusters of size 40, and the right panels showcase 40 clusters of size 10. The CPs of the 95\% CIs for most of the R packages were found to be very close to 95\%. \textcolor{red}{However, in the scenario of 10 clusters of size 10, \texttt{frailtypack} yielded  slightly lower CP for $\beta_1$ and $\beta_2$. Similarly, in the case of 10 clusters of size 40, \texttt{frailtypack} yielded slightly lower CP for $\beta_2$, while \texttt{frailtySurv} had slightly lower CP for $\beta_3$.} The detailed results are displayed in Table \ref{cptable} in the Appendix. 

For the variance of the frailty terms, most of the R packages calculate the CI  based on normal approximation as $\hat{\theta} \pm 1.96* \SE(\hat{\theta})$. We name this type of interval as CI$^{(1)}$.  The fourth row in Figure \ref{fig: cp}  clearly showed the CPs  for CI$^{(1)}$ failed to attain the 95\%  nominal level. \textcolor{red}{Moreover, in the scenario of 10 clusters of size 10 with an 80\% censoring rate, CPs  for CI$^{(1)}$ exceeded the 95\% nominal level in the parfm, frailtyEM, and frailtySurv packages. These results are not surprising, since the distribution of the variance of the frailty terms is widely known for being skewed as shown by Figure \ref{fig: random_effect}; this was previously reported by \cite{McCullochCharlesE.2011MtSo}.} Better CI may be constructed with the sampling distribution of the logarithm of the variance of the frailty terms, which is more symmetric \cite{balan_ascertainment_2016}. Then, the 95\% CI for $\log \hat{\theta}$  can be constructed as 
\begin{equation}
 [\log \hat{\theta} - 1.96\times\SE(\log \hat{\theta}), \log \hat{\theta} + 1.96\times \SE(\log \hat{\theta})].
\end{equation}
The  95\% CI for $\hat{\theta}$ can be then calculated by exponentiating the lower and upper boundaries of the 95\% CI for $\log \hat{\theta}$. We call this type of interval CI$^{(2)}$. Most R packages do not provide the value of $\SE(\log \hat{\theta})$ directly. However, we can calculate it from the $\SE(\hat\theta)$ using the relationship of the Fisher information \cite{GatenbyRobertA.2013TCRo} between $\theta$ and its log transformation $\phi=\log(\theta)$, which is derived briefly in general terms as follows. Let $X$ be a random vector (data) with the PDF $f(x\mid\theta)$.  Let $I_1(\theta)$ denote the Fisher information of $\theta$ and $l_1(\theta;x)$ denote the log-likelihood of $\theta$ given $x$. Suppose we re-parameterize $\theta = \Theta(\phi)$, where $\Theta(\cdot)$ is a differentiable function. The log-likelihood function for $\phi$, $l_2(\phi;x)$, is given by: 
\begin{equation}
l_2(\phi;x) = l_1(\Theta(\phi);x)=\log f(x\mid \Theta(\phi)). 
\end{equation}
Then the derivative of $l_2$ is given by:
\begin{equation}
\frac{\partial l_2(\phi;x)}{\partial \phi}= \frac{\frac{\partial f(x\mid \theta)}{\partial \theta}\frac{\partial \Theta(\phi)}{\partial \phi}}{f(x\mid\Theta(\phi))}.
\end{equation}
It follows that the Fisher's information of $\phi$, $I_2(\phi)$, is obtained as follows:
\begin{equation}
I_2(\phi) =E_X\left\{\left(\frac{\frac{\partial f(X\mid \theta)}{\partial \theta}}{f(X\mid\theta(\phi))} \right)^2\right \} \left(\frac{\partial \Theta(\phi)}{\partial \phi} \right)^2=I_1(\theta)(\Theta'(\phi) )^2,
\end{equation}
where $\Theta'$ denotes the derivative function of $\Theta$. Applying the above general rule to $\Theta(\phi)=\exp(\phi)$ (ie, $\phi=\log(\theta)$), we arrive at the following equation:
\begin{equation}
I_2(\phi) = I_1(\theta) \theta^2. \label{eqn:fish}
\end{equation}
We know that $\SE(\hat{\theta}) =\frac{1}{\sqrt{I_1(\theta)}}$, where $\hat{\theta}$ is the maximum likelihood estimation (MLE) of $\theta$. Finally, we arrive at the following relationship:
\begin{equation}
\SE(\log(\hat \theta))=\frac{1}{\sqrt{I_2(\phi)}} =\frac{1}{\theta} \frac{1}{\sqrt{I_1(\theta)}} =\frac{1}{\theta} \SE(\hat{\theta}).
\end{equation}
\noindent

As shown in the fifth row of the Figure \ref{fig: cp}, CI$^{(2)}$ had consistently higher CP than  CI$^{(1)}$ across most packages. \textcolor{red}{However, the CI$^{(2)}$ exhibited coverage probabilities exceeding the 95\% nominal level in the frailtypack package for the scenarios with an all censoring rate of 100 sample size and an 80\% censoring of 400 sample size.} Interestingly, as the censoring rate increases, CPs of both CI$^{(1)}$ and CI$^{(2)}$ for $\hat{\theta}$ became  closer to 95\%. This is partly due to  the larger variability of the estimated parameters as a result of  the higher censoring. This finding is consistent with the results of Balan et al. \cite{balan_ascertainment_2016}. In addition, all packages had lower CPs in the scenario with 10 clusters compared to the scenario with 40 clusters; this is presumably caused by the shape of the sampling distribution of the variance of the frailty terms (or its log) being closer to normal when the number of clusters is larger.

\clearpage
\subsection{Convergence rate}

Table \ref{tab: crtable} presents the results of the convergence rate  of each package. The \texttt{survival}, \texttt{frailtySurv}, and  \texttt{frailtypack} packages had convergence rates over  97\%  in all scenarios. When the sample size is small with a large censoring rate, \texttt{frailtyHL}, \texttt{parfm} and \texttt{frailtyEM} packages had relatively lower convergence rates.  In the scenario with an extremely low sample size of 100 at 80\% censorship, the \texttt{parfm} and \texttt{frailtyEM} packages had the lowest convergence rate compared with other packages at about 66.3\% and 57.2\%, respectively.

\begin{table}[htb]
\tiny
\centering
\caption{Convergence rate of the R packages over 1000 simulated datasets. Note that some very poorly fitted models are considered as not convergence.}
\begin{tabular}{llllllllll}
\hline
$n$&{Clusters}&{Obs} &$100c$ &{survival} &{parfm}&{frailtyEM} &{frailtySurv}&{frailtyHL}&{frailtypack}\\
  \hline
  100 &10 &10 &20 & 100 & 96.3 & 93.3 & 99.8 & 97.9 & 99.4  \\ 
  400 &40 &10 &20 & 100 & 100 & 99.1 & 100 & 100 & 99.6  \\ 
  400 &10 &40 &20 & 100 & 99.9 & 96.2 & 100 & 99.2 & 99.3 \\ 
  800 &80 &10 &20 & 100 & 100 & 99.4 & 100 & 100 & 99.8  \\ 
  800 &10 &80 &20 & 100 & 100 & 94.7 & 100 & 98.7 & 99.5  \\ 
  \hline
  100 &10 &10 &50 & 99.9 & \textbf{88.5} & \textbf{83.1} & 99.2 & 97.1 & 100  \\ 
  400 &40 &10 &50 & 100 & 100 & 99.6 & 100 & 100 & 99.7  \\ 
  400 &10 &40 &50 & 100 & 99.5 & 97.5 & 100 & 99.2 & 99.4 \\
  800 &80 &10 &50 & 100 & 100 & 99.5 & 100 & 100 & 100 \\ 
  800 &10 &80 &50 & 100 & 100 & 95.8 & 100 & 98.7 & 98.5 \\ 
  \hline
  100 &10 &10 &80 & 97.2 & \textbf{66.3} & \textbf{57.2} & 98.1 & \textbf{91.4} & 99.1 \\ 
  400 &40 &10 &80 & 100 & 97.7 & 95.8 & 100 & 99.5 & 99.7  \\ 
  400 &10 &40 &80 & 100 & 96 & 93.2 & 99.7 & 98.1 & 99.8  \\ 
  800 &80 &10 &80 & 100 & 99.9 & 99.3 & 100 & 99.9 & 99.8 \\ 
  800 &10 &80 &80 & 100 & 99.5 & 97.8 & 100 & 99.2 & 99.8 \\ 
  \hline
\end{tabular}
\label{tab: crtable}
\end{table}

\subsection{Computing time}
Table \ref{tab: timetable} reports the average computing time for fitting the shared frailty model using the R packages under each simulation scenario. The package \texttt{survival} is the fastest one, followed by \texttt{frailtyEM} and \texttt{frailtypack}, and \texttt{parfm}, \texttt{frailtySurv} and \texttt{frailtyHL}. In general, the larger the number of clusters and cluster size  requires more computing time for most packages, except for \texttt{frailtyEM} package.

\begin{table}[htb]
\tiny
\centering
\caption{Average computing time (\textcolor{red}{in seconds}) of the R packages under each simulation scenario. }
\begin{tabular}{llllllllll}
\hline
$n$&{Clusters}&{Obs} &$100c$ &{survival} &{parfm}&{frailtyEM} &{frailtySurv}&{frailtyHL}&{frailtypack}\\
  \hline
  100 &10 &10 &20 & 0.013 & 3.551  & 0.486 & 0.248   & 0.463   & 0.451 \\ 
  400 &40 &10 &20 & 0.030 & 10.693 & 1.319 & 14.579  & 19.694  & 1.135 \\ 
  400 &10 &40 &20 & 0.020 & 6.581  & 2.151 & 9.881   & 5.044   & 1.097 \\ 
  800 &80 &10 &20 & 0.059 & 17.854 & 2.310 & 164.660 & 123.478 & 2.222 \\ 
  800 &10 &80 &20 & 0.032 & 15.908 & 6.512 & 124.758 & 44.476  & 2.239 \\ 
  \hline
  100 &10 &10 &50 & 0.014 & 3.859   & 0.388  & 0.233   & 0.553  & 0.409 \\ 
  400 &40 &10 &50 & 0.020 & 7.281   & 0.763  & 13.860  & 11.572 & 1.022  \\ 
  400 &10 &40 &50 & 0.020 & 7.281   & 0.763 & 13.860  & 11.572  & 1.022 \\
  800 &80 &10 &50 & 0.059 & 24.918  & 2.068 & 230.582 & 160.960 & 2.450 \\ 
  800 &10 &80 &50 & 0.023 & 10.093  & 3.450 & 125.591 & 19.418  & 1.656 \\ 
  \hline
  100 &10 &10 &80 & 0.011  & 3.180  & 0.278 & 0.184   & 0.643   & 0.266\\ 
  400 &40 &10 &80 & 0.021  & 9.575  & 0.478 & 13.555  & 22.614  & 0.860  \\ 
  400 &10 &40 &80 & 0.017  & 6.551  & 0.611 & 9.659   & 7.316   & 0.902 \\ 
  800 &80 &10 &80 & 0.032  & 16.223 & 0.754 & 162.441 & 141.127 & 1.394 \\ 
  800 &10 &80 &80 & 0.020  & 9.521  & 1.412 & 124.499 & 18.479  & 1.235 \\ 
  \hline
\end{tabular}
\label{tab: timetable}
\end{table}

\section{Conclusions and Discussions}\label{sec5}
In this paper, all the R packages considered for fitting the shared frailty models gave very similar and unbiased parameter estimates for the fixed-effect regression coefficients, regardless of the sample size, cluster sizes and  censoring rates. However, there were differences between the packages with respect to the estimation of the variance parameter for the frailty term. In general,  the variance parameter of the frailty term was consistently underestimated for all the R packages considered in this paper. However, as the  censoring rate increases, the bias is less pronounced but subject to more variability, which leads to higher MSE. This finding  is consistent with the finding in other studies \cite{HirschKatharina2011Sfss, petersen_inference_2006}. Our results also showed that a larger number of clusters can lead to a higher precision of the estimated variance parameter of the frailty term. The CP of the 95\% CIs of the regression coefficients for most of the R packages are very close to 95\%, and all packages of the variance of the frailty terms had lower CP in the scenario with  a smaller number of clusters compared to the scenario with  a larger  number of clusters. Most packages had convergence rates over  97\%  in all scenarios, except for the \texttt{parfm} and \texttt{frailtyEM} packages in the scenario with a small sample size (n=100) and large censorship (80\%). The computing time for all scenarios of \texttt{survival}, \texttt{frailtyEM} and \texttt{frailtypack} packages are within 0.1 minutes; the \texttt{parfm} takes no more than 0.5 minutes. However, the computing time for \texttt{frailtySurv} and \texttt{frailtyHL} packages need two to three minutes under the sample size $n$=800.

The best package to estimate the parameters of a frailty model is  the  \texttt{survival} package, which is computationally fast with a high convergence rate in almost all simulation scenarios. However, the \texttt{survival} package does not provide the estimate of standard error for the variance component of the frailty. Since the EM and PPL algorithms lead to the same estimates  in a frailty model, \texttt{frailtyEM} can be used to substitute survival if the standard error of the variance of the frailty terms is required in a real application. However, we do not suggest using \texttt{frailtyEM} package when  the sample size is small with a large censoring rate due to its lower convergence rate. The \texttt{parfm} has a lower convergence rate as well in the scenario with a small sample size at a large censoring rate. The parametric estimation is more powerful if the baseline hazard distribution is known, then the \texttt{parfm} is a good choice in the large sample size study. The \texttt{frailtySurv} fits the frailty model with a wide range of frailty distributions, and the \texttt{frailtyHL} allows multilevel frailties in the frailty model. \textcolor{red}{However, the \texttt{parfm}, \texttt{frailtySurv} and \texttt{frailtyHL} packages require more computing time compared to other methods,  which is due to the method used for modelling the frailty term. When the sample size is large, the \texttt{frailtypack} package demonstrates similar performance in parameter estimation to other packages. However, it may produce less accurate estimates when the sample size is 100. Hence, caution should be exercised when using the \texttt{frailtypack} package for datasets with small sample sizes and few clusters. To improve the performance of \texttt{frailtypack} package, we may need carefully choose the number of knots and other parameters for using spline methods. The number of knots may play a crucial role in estimating the baseline hazard using spline methods. The baseline hazard function might impact the estimated variance of the frailty term. Early research \cite{HirschKatharina2011Sfss} also noted that using a smaller number of knots typically helps to circumvent the problem of overestimating the fixed effect. On the other hand, the \texttt{frailtypack} package offers the advantage of accommodating more complex structures for frailty terms. This includes nested and joined frailties, as well as frailty interactions, enabling more flexible modelling options.}

In this \textcolor{red}{paper}, a new type of confidence interval for the variance of the frailty terms $\theta$, using the standard error of $\log(\hat{\theta})$ was implemented. The coverage probability of the proposed confidence interval is much higher than the confidence interval based on the standard error of the variance of the frailty terms. Most packages do not provide the standard error of $\log \hat\theta$. Our proposed approach provides a solution by using the Fisher information approach. We recommend adding this approach to the R packages for calculating a more reliable 95\% confidence interval for the variance of the frailty terms in frailty models.















\newpage
 \begin{appendices}

\section{Additional Tables}\label{secA1}





\end{appendices}

\begin{table}[htp]
\tiny
\centering
\caption{Performance of the parameter estimation of different R packages. We only considered the converged fitted models for 1000 simulated datasets for each package. The total sample size is 100 and the censorship is 20\%.}

\begin{tabular}{lllllllll}
\hline
 Parameter &True &Mean&Bias& Mean.se  & Emp. se & Median & Median.se& MSE\\
  \hline
  survival& & & & & & & & \\
 
  $\beta_1$ & 1.000 & 1.014 & 0.014 & 0.447 & 0.459 & 1.004 & 0.444& 0.459\\ 
  $\beta_2$ & -1.000 & -1.011 & -0.011 & 0.158 & 0.172 & -1.004 & 0.157& 0.172\\ 
  $\beta_3$ & 0.500 & 0.508 & 0.008 & 0.292 & 0.309 & 0.493 & 0.289 & 0.309\\ 
  $\theta$ & 0.500 & 0.404 & -0.096 & - & 0.353 & 0.318 & -& 0.362\\ 
  \hline
  parfm& & & & & & & \\
  $\beta_1$ & 1.000 & 1.027 & 0.027 & 0.436 & 0.452 & 1.018 & 0.434 & 0.453\\ 
  $\beta_2$ & -1.000 & -1.028 & -0.028 & 0.156 & 0.166 & -1.018 & 0.155&0.167 \\ 
  $\beta_3$ & 0.500 & 0.515 & 0.015 & 0.288 & 0.303 & 0.511 & 0.286 & 0.303\\ 
  $\theta$ & 0.500 & 0.427 & -0.073 & 0.261 & 0.401 & 0.329 & 0.219& 0.406\\ 
  
  \hline
  frailtyEM& & & & &  & & \\
  $\beta_1$ & 1.000 & 1.014 & 0.014 & 0.447 & 0.449 & 1.003 & 0.443 & 0.449\\ 
  $\beta_2$ & -1.000 & -1.013 & -0.013 & 0.158 & 0.167 & -1.005 & 0.157& 0.167\\ 
  $\beta_3$ & 0.500 & 0.508 & 0.008 & 0.291 & 0.300 & 0.492 & 0.288 & 0.300\\ 
    $\theta$& 0.500 & 0.401 & -0.099 & 0.255 & 0.348 & 0.320 & 0.217 & 0.358\\ 
  
  \hline
   frailtySurv& & & & & & & \\
   $\beta_1$ & 1.000 & 1.034 & 0.034 & 0.530 & 0.459 & 1.025 & 0.502 & 0.460 \\ 
  $\beta_2$ & -1.000 & -1.017 & -0.017 & 0.185 & 0.172 & -1.006 & 0.173 & 0.172\\ 
  $\beta_3$ & 0.500 & 0.518 & 0.018 & 0.288 & 0.311 & 0.500 & 0.273 &0.311\\ 
   $\theta$ & 0.500 & 0.385 & -0.115 & 0.273 & 0.356 & 0.299 & 0.197 &0.369\\
   
   \hline
  frailtyHL& & & & & & & \\
  $\beta_1$ & 1.000 & 1.032 & 0.032 & 0.449 & 0.463 & 1.021 & 0.445& 0.464\\ 
  $\beta_2$ & -1.000 & -1.029 & -0.029 & 0.160 & 0.174 & -1.023 & 0.158& 0.175\\ 
  $\beta_3$ & 0.500 & 0.517 & 0.017 & 0.293 & 0.310 & 0.499 & 0.290& 0.310\\ 
   $\theta$ & 0.500 & 0.461 & -0.039 & 0.276 & 0.362 & 0.379 & 0.242& 0.364\\
  
   \hline
   frailtypack& & & & & & & \\
  $\beta_1$ & 1.000 & 0.992 & -0.008 & 0.397 & 0.456 & 0.951 & 0.419 & 0.456\\ 
  $\beta_2$ & -1.000 & -0.874 & 0.126 & 0.145 & 0.197 & -0.876 & 0.146 &0.213\\ 
  $\beta_3$ & 0.500 & 0.467 & -0.033 & 0.280 & 0.289 & 0.454 & 0.278 &0.290\\ 
   $\theta$ & 0.500 & 0.253 & -0.247 & 0.176 & 0.262 & 0.182 & 0.150& 0.323 \\
  
  \hline
\end{tabular}
\label{table_100_20}
\end{table}


\begin{table}[htp]
\tiny
\centering
\caption{Performance of the parameter estimation of different R packages. We only considered the converged fitted models for 1000 simulated datasets for each package. The total sample size is 100 and the censorship is 50\%.}

\begin{tabular}{lllllllll}
\hline
 Parameter &True &Mean&Bias& Mean.se  & Emp. se & Median & Median.se& MSE\\
  \hline
  survival& & & & & & & & \\

  $\beta_1$ & 1.000 & 1.016 & 0.016 & 0.598 & 0.623 & 1.013 & 0.591& 0.623 \\ 
  $\beta_2$ & -1.000 & -1.025 & -0.025 & 0.207 & 0.231 & -1.014 & 0.203& 0.232 \\ 
  $\beta_3$ & 0.500 & 0.504 & 0.004 & 0.386 & 0.412 & 0.490 & 0.379& 0.412 \\ 
  $\theta$ & 0.500 & 0.423 & -0.077 & - & 0.353 & 0.329 &  -& 0.359\\ 
      
  \hline
  parfm& & & & & & & \\
  $\beta_1$ & 1.000 & 1.031 & 0.031 & 0.584 & 0.577 & 1.022 & 0.579& 0.578 \\ 
  $\beta_2$ & -1.000 & -1.053 & -0.053 & 0.207 & 0.210 & -1.047 & 0.204 & 0.213\\ 
  $\beta_3$ & 0.500 & 0.517 & 0.017 & 0.385 & 0.394 & 0.494 & 0.378& 0.394 \\ 
  $\theta$ & 0.500 & 0.449 & -0.051 & 0.320 & 0.401 & 0.346 & 0.278& 0.404 \\ 
  
  \hline
  frailtyEM& & & & &  & & \\
   $\beta_1$ & 1.000 & 1.022 & 0.022 & 0.599 & 0.576 & 1.014 & 0.593& 0.576 \\ 
  $\beta_2$ & -1.000 & -1.028 & -0.028 & 0.208 & 0.208 & -1.015 & 0.203& 0.209 \\ 
  $\beta_3$ & 0.500 & 0.499 & -0.001 & 0.388 & 0.382 & 0.474 & 0.381& 0.382 \\ 
  $\theta$ & 0.500 & 0.424 & -0.076 & 0.316 & 0.348 & 0.332 & 0.279& 0.354 \\ 
  
  \hline
   frailtySurv& & & & & & & \\
   $\beta_1$ & 1.000 & 1.030 & 0.030 & 0.903 & 0.618 & 1.003 & 0.729& 0.619 \\ 
  $\beta_2$ & -1.000 & -1.031 & -0.031 & 0.274 & 0.230 & -1.018 & 0.224& 0.231 \\ 
  $\beta_3$ & 0.500 & 0.516 & 0.016 & 0.428 & 0.413 & 0.498 & 0.366 & 0.413\\ 
  $\theta$ & 0.500 & 0.364 & -0.136 & 0.478 & 0.361 & 0.268 & 0.280& 0.379 \\ 
   
   \hline
  frailtyHL& & & & & & & \\
  $\beta_1$ & 1.000 & 1.042 & 0.042 & 0.605 & 0.633 & 1.030 & 0.599& 0.635 \\ 
  $\beta_2$ & -1.000 & -1.058 & -0.058 & 0.211 & 0.236 & -1.045 & 0.207& 0.239 \\ 
  $\beta_3$ & 0.500 & 0.520 & 0.020 & 0.391 & 0.420 & 0.504 & 0.384& 0.420 \\ 
  $\theta$ & 0.500 & 0.489 & -0.011 & 0.332 & 0.401 & 0.403 & 0.300 & 0.401\\ 
  
   \hline
   frailtypack& & & & & & & \\
 $\beta_1$ & 1.000 & 1.060 & 0.060 & 0.518 & 0.639 & 0.998 & 0.540& 0.643 \\ 
  $\beta_2$ & -1.000 & -0.903 & 0.097 & 0.193 & 0.243 & -0.860 & 0.190& 0.252 \\ 
  $\beta_3$ & 0.500 & 0.482 & -0.018 & 0.369 & 0.385 & 0.458 & 0.361 &0.385 \\ 
  $\theta$ & 0.500 & 0.237 & -0.263 & 0.196 & 0.273 & 0.155 & 0.183& 0.342 \\

  \hline
\end{tabular}
\label{table_100_50}

\end{table}

\newpage

\begin{table}[htp]
\tiny
\centering
\caption{Performance of the parameter estimation of different R packages. We only considered the converged fitted models for 1000 simulated datasets for each package. The total sample size is 100 and the censorship is 80\%.}

\begin{tabular}{lllllllll}
\hline
 Parameter &True &Mean&Bias& Mean.se  & Emp. se & Median & Median.se& MSE\\
  \hline
  survival& & & & & & & & \\
  $\beta_1$  & 1.000 & 1.126 & 0.126 & 1.028 & 1.100 & 1.058 & 0.990&1.116 \\ 
  $\beta_2$  & -1.000 & -1.083 & -0.083 & 0.349 & 0.407 & -1.031 & 0.329&0.414 \\ 
  $\beta_3$  & 0.500 & 0.496 & -0.004 & 0.656 & 0.710 & 0.494 & 0.624& 0.710\\ 
  $\theta$ & 0.500 & 0.647 & 0.147 &  & 0.475 & 0.470 & - & 0.497 \\ 
      
  \hline
  parfm& & & & & & & \\
  $\beta_1$  & 1.000 & 1.172 & 0.172 & 1.018 & 0.877 & 1.108 & 0.975&0.907 \\ 
  $\beta_2$  & -1.000 & -1.130 & -0.130 & 0.364 & 0.342 & -1.063 & 0.338&0.359 \\ 
  $\beta_3$  & 0.500 & 0.552 & 0.052 & 0.675 & 0.618 & 0.557 & 0.645 & 0.621\\ 
  $\theta$ & 0.500 & 0.819 & 0.319 & 0.743 & 0.834 & 0.547 & 0.592& 0.936\\ 
  
  \hline
  frailtyEM& & & & &  & & \\
   $\beta_1$  & 1.000 & 1.237 & 0.237 & 1.042 & 0.844 & 1.169 & 1.003 & 0.900\\ 
  $\beta_2$  & -1.000 & -1.106 & -0.106 & 0.355 & 0.319 & -1.055 & 0.336&0.330 \\ 
  $\beta_3$  & 0.500 & 0.499 & -0.001 & 0.661 & 0.571 & 0.527 & 0.633 & 0.571\\ 
  $\theta$ & 0.500 & 0.700 & 0.200 & 0.710 & 0.531 & 0.491 & 0.591& 0.571\\ 
  
  \hline
   frailtySurv& & & & & & & \\
   $\beta_1$  & 1.000 & 1.140 & 0.140 & 3.493 & 1.078 & 1.086 & 1.483& 1.098 \\ 
  $\beta_2$  & -1.000 & -1.091 & -0.091 & 0.930 & 0.412 & -1.036 & 0.399&0.420 \\ 
  $\beta_3$  & 0.500 & 0.504 & 0.004 & 1.839 & 0.710 & 0.506 & 0.695&0.710 \\ 
  $\theta$ & 0.500 & 0.407 & -0.093 & 2.031 & 0.623 & 0.144 & 0.690 & 0.632\\ 
   
   \hline
  frailtyHL& & & & & & & \\
   $\beta_1$  & 1.000 & 1.244 & 0.244 & 1.083 & 1.152 & 1.109 & 1.032&1.212 \\ 
  $\beta_2$  & -1.000 & -1.191 & -0.191 & 0.373 & 0.460 & -1.104 & 0.350&0.497 \\ 
  $\beta_3$  & 0.500 & 0.550 & 0.050 & 0.691 & 0.782 & 0.520 & 0.657&0.785 \\ 
  $\theta$ & 0.500 & 0.825 & 0.325 & 0.739 & 0.837 & 0.632 & 0.623 & 0.943\\ 
  
   \hline
   frailtypack& & & & & & & \\
  $\beta_1$  & 1.000 & 1.602 & 0.602 & 0.938 & 1.319 & 1.404 & 0.907&1.681 \\ 
  $\beta_2$  & -1.000 & -1.081 & -0.081 & 0.351 & 0.471 & -0.997 & 0.319& 0.478\\ 
  $\beta_3$  & 0.500 & 0.586 & 0.086 & 0.643 & 0.758 & 0.560 & 0.609& 0.765 \\ 
  $\theta$ & 0.500 & 0.376 & -0.124 & 0.375 & 0.664 & 0.091 & 0.325 & 0.679\\   
  \hline
\end{tabular}
\label{table_100_80}

\end{table}

\newpage

\begin{table}[htb]
\tiny
\centering
\caption{Performance of the parameter estimation of different R packages. We only considered the converged fitted models for 1000 simulated datasets for each package. The total sample size is 400 and the censorship is 20\%.}
\begin{tabular}{lllllllll}
\hline
 Parameter &True &Mean&Bias& Mean.se  & Emp. se & Median&Median.se& MSE\\

 \hline
 10 clusters of size 40\\
  \hline
  survival& & & & & & & & \\
  $\beta_1$  & 1.000 & 0.996 & -0.004 & 0.208 & 0.198 & 0.995 & 0.208 &0.198 \\ 
  $\beta_2$  & -1.000 & -1.004 & -0.004 & 0.074 & 0.073 & -1.001 & 0.074&0.073 \\ 
  $\beta_3$  & 0.500 & 0.501 & 0.001 & 0.136 & 0.133 & 0.503 & 0.136& 0.133\\ 
  $\theta$ & 0.500 & 0.568 & 0.068 & - & 0.382 & 0.510 & -  &0.387 \\ 

  \hline
  parfm& & & & & & & & \\
   $\beta_1$  & 1.000 & 0.993 & -0.007 & 0.205 & 0.197 & 0.995 & 0.204& 0.197\\ 
  $\beta_2$  & -1.000 & -1.002 & -0.002 & 0.072 & 0.072 & -1.001 & 0.072&0.072 \\ 
  $\beta_3$  & 0.500 & 0.500 & 0.000 & 0.134 & 0.133 & 0.499 & 0.134& 0.133\\ 
  $\theta$ & 0.500 & 0.468 & -0.032 & 0.211 & 0.382 & 0.357 & 0.169&0.383 \\ 
  \hline
  frailtyEM& & & & &  & & & \\
   $\beta_1$  & 1.000 & 0.991 & -0.009 & 0.207 & 0.193 & 0.990 & 0.207& 0.193\\ 
  $\beta_2$  & -1.000 & -0.999 & 0.001 & 0.074 & 0.071 & -0.997 & 0.074& 0.071\\ 
  $\beta_3$  & 0.500 & 0.500 & 0.000 & 0.135 & 0.130 & 0.502 & 0.135&0.130 \\ 
  $\theta$ & 0.500 & 0.462 & -0.038 & 0.206 & 0.372 & 0.346 & 0.164& 0.373\\ 
  
  \hline
   frailtySurv& & & & & & & & \\
   $\beta_1$  & 1.000 & 1.000 & 0.000 & 0.209 & 0.205 & 0.998 & 0.207& 0.205\\ 
  $\beta_2$  & -1.000 & -0.999 & 0.001 & 0.086 & 0.079 & -0.998 & 0.084& 0.079\\ 
  $\beta_3$  & 0.500 & 0.503 & 0.003 & 0.129 & 0.138 & 0.502 & 0.127& 0.138\\ 
  $\theta$ & 0.500 & 0.567 & 0.067 & 0.308 & 0.391 & 0.481 & 0.171& 0.395\\ 
   
   \hline
  frailtyHL& & & & & & & & \\
  $\beta_1$  & 1.000 & 0.995 & -0.005 & 0.207 & 0.197 & 0.994 & 0.207& 0.197\\ 
  $\beta_2$  & -1.000 & -1.002 & -0.002 & 0.074 & 0.072 & -1.000 & 0.074& 0.072\\ 
  $\beta_3$  & 0.500 & 0.501 & 0.001 & 0.135 & 0.133 & 0.502 & 0.135&0.133 \\ 
  $\theta$ & 0.500 & 0.516 & 0.016 & 0.237 & 0.404 & 0.394 & 0.192& 0.404\\ 
  
   \hline
   frailtypack& & & & & & & & \\
 $\beta_1$  & 1.000 & 1.016 & 0.016 & 0.206 & 0.198 & 1.017 & 0.206& 0.198\\ 
  $\beta_2$  & -1.000 & -0.984 & 0.016 & 0.073 & 0.078 & -0.985 & 0.073& 0.078\\ 
  $\beta_3$  & 0.500 & 0.500 & 0.000 & 0.135 & 0.133 & 0.503 & 0.134& 0.133\\ 
  $\theta$ & 0.500 & 0.480 & -0.020 & 0.227 & 0.475 & 0.343 & 0.167& 0.475\\

  \hline
  40 clusters of size 10 \\
  \hline
  survival& & & & & & & & \\
  $\beta_1$  & 1.000 & 0.992 & -0.008 & 0.217 & 0.217 & 0.993 & 0.216& 0.217\\ 
  $\beta_2$  & -1.000 & -0.990 & 0.010 & 0.076 & 0.081 & -0.991 & 0.076&0.081 \\ 
  $\beta_3$  & 0.500 & 0.488 & -0.012 & 0.142 & 0.139 & 0.486 & 0.142&0.139 \\ 
  $\theta$  & 0.500 & 0.389 & -0.111 & -& 0.163 & 0.366 & - &0.175 \\ 
  
  \hline
  parfm& & & & & & & & \\
  $\beta_1$  & 1.000 & 0.991 & -0.009 & 0.215 & 0.214 & 0.990 & 0.214& 0.214 \\ 
  $\beta_2$  & -1.000 & -0.989 & 0.011 & 0.076 & 0.080 & -0.988 & 0.076 &0.080 \\ 
  $\beta_3$  & 0.500 & 0.488 & -0.012 & 0.141 & 0.139 & 0.487 & 0.141 & 0.139\\ 
  $\theta$ & 0.500 & 0.391 & -0.109 & 0.127 & 0.162 & 0.367 & 0.121&0.174 \\ 
  
  \hline
  frailtyEM& & & & &  & & & \\
  $\beta_1$  & 1.000 & 0.991 & -0.009 & 0.217 & 0.215 & 0.993 & 0.216 &0.215 \\ 
  $\beta_2$  & -1.000 & -0.989 & 0.011 & 0.076 & 0.081 & -0.990 & 0.076 & 0.081\\ 
  $\beta_3$  & 0.500 & 0.488 & -0.012 & 0.142 & 0.138 & 0.486 & 0.142 & 0.138\\ 
  $\theta$ & 0.500 & 0.386 & -0.114 & 0.126 & 0.157 & 0.365 & 0.121 & 0.170\\ 
  
  \hline
   frailtySurv& & & & & & & &  \\
   $\beta_1$  & 1.000 & 0.991 & -0.009 & 0.249 & 0.223 & 0.990 & 0.244&0.223 \\ 
  $\beta_2$  & -1.000 & -0.986 & 0.014 & 0.098 & 0.095 & -0.993 & 0.090& 0.095\\ 
  $\beta_3$  & 0.500 & 0.492 & -0.008 & 0.144 & 0.143 & 0.490 & 0.141&0.143  \\ 
  $\theta$ & 0.500 & 0.438 & -0.062 & 0.408 & 0.223 & 0.392 & 0.133& 0.227\\ 
   
   \hline
  frailtyHL& & & & & & & & \\
  $\beta_1$  & 1.000 & 0.996 & -0.004 & 0.217 & 0.218 & 0.996 & 0.216&0.218 \\ 
  $\beta_2$  & -1.000 & -0.994 & 0.006 & 0.076 & 0.081 & -0.995 & 0.076&0.081 \\ 
  $\beta_3$  & 0.500 & 0.490 & -0.010 & 0.142 & 0.140 & 0.488 & 0.142&0.140 \\ 
  $\theta$ & 0.500 & 0.411 & -0.089 & 0.129 & 0.164 & 0.388 & 0.124&0.172 \\ 
  
   \hline
   frailtypack& & & & & & & & \\
  $\beta_1$  & 1.000 & 1.014 & 0.014 & 0.215 & 0.217 & 1.013 & 0.216&0.217 \\ 
  $\beta_2$  & -1.000 & -0.968 & 0.032 & 0.076 & 0.095 & -0.974 & 0.076&0.096 \\ 
  $\beta_3$  & 0.500 & 0.488 & -0.012 & 0.141 & 0.139 & 0.486 & 0.141& 0.139\\ 
  $\theta$ & 0.500 & 0.359 & -0.141 & 0.119 & 0.149 & 0.343 & 0.116& 0.169\\ 
 \hline

\end{tabular}
\label{table1}

\end{table}
 
\newpage

\begin{table}[htb]
\tiny
\centering
\caption{Performance of the parameter estimation of different R packages. We only considered the converged fitted models for 1000 simulated datasets for each package. The total sample size is 400 and the censorship is 50\%.}
\begin{tabular}{lllllllll}
\hline
 Parameter &True &Mean&Bias& Mean.se  & Emp. se & Median&Median.se& MSE\\
\hline
   40 clusters of size 10 \\
  \hline
  survival& & & & & & & & \\
  $\beta_1$ &1&0.993&	-0.007&	0.283&	0.29&	0.984&	0.283&0.290 \\ 
  $\beta_2$ &-1&-0.99&	0.01&	0.097&	0.102&	-0.992&	0.096& 0.102\\
  $\beta_3$ & 0.5&0.486	& -0.014&	0.183&	0.182&	0.489&	0.182 & 0.182\\
  $\theta$&0.5&0.383	&-0.117& - &	0.183&	0.356&	-& 0.197\\
 \hline
  parfm& & & & & & & & \\
  $\beta_1$ &1&0.997&	-0.003&	0.28&	0.287&	0.989&	0.279&0.287 \\
  $\beta_2$ &-1&-0.993& 0.007&	0.097&	0.1&  -0.994&	0.097&0.100 \\
  $\beta_3$ &0.5&0.487	&-0.013&	0.182&	0.18& 0.492& 0.181&0.180 \\
  $\theta$&0.5&0.395	&-0.105&	0.147&	0.18 &  0.364& 0.141&0.191 \\
  \hline
  frailtyEM& & & & &  & & &  \\
  $\beta_1$ &1&0.994&-0.006&	0.283&	0.289&	0.985&	0.282&0.289 \\
  $\beta_2$ & -1&-0.991&0.009&	0.097&	0.102&	-0.992&	0.096&0.102 \\
  $\beta_3$ &0.5&0.487	&-0.013&	0.183&	0.181&	0.49 & 0.182&0.181 \\
  $\theta$&0.5&0.388	&-0.112&	0.147&	0.177&	0.361&	0.141&0.190 \\
  \hline
  frailtySurv& & & & & & & & \\
  $\beta_1$ &1&0.990&-0.01&	0.337&	0.29 &  0.988&	0.336&0.290 \\
  $\beta_2$ &-1&-0.997&0.003& 0.11& 0.105&	-0.997&	0.108& 0.105 \\
  $\beta_3$ & 0.5&0.493	&-0.007&	0.184&	0.186&	0.497&	0.181&0.186 \\
  $\theta$&0.5&0.396	&-0.104&	0.164&	0.182&	0.374&	0.149& 0.193\\
  \hline
  frailtyHL& & & & & & & & \\
  $\beta_1$ & 1&1.002&0.002	&  0.284&	0.292&	0.996&	0.284& 0.292 \\
  $\beta_2$ & -1&-0.999&0.001&	0.097&	0.102&	-1 &  0.097&0.102 \\
  $\beta_3$ &0.5&0.49&-0.01	&      0.184&	0.183&	0.493&	0.183& 0.183\\
  $\theta$&0.5&0.424	&-0.076&	0.152&	0.185&	0.395&	0.145 &0.191  \\
  \hline
   frailtypack& & & & & & & & \\
  $\beta_1$ &1   &1.018     &0.018	& 0.274  &  0.295&	1.01& 0.279&0.295 \\
  $\beta_2$ &-1   &-0.953  &0.047    &0.096 &	0.126&    -0.967&	0.097&0.128 \\
  $\beta_3$ &0.5 &0.481	&-0.019	&0.181&	0.177&     0.476&	0.181& 0.177\\
  $\theta$&0.5 &0.342	&-0.158	&0.133&	0.171&	0.322&	0.131&0.195 \\
  \hline
 10 clusters of size 40\\
  \hline
  survival& & & & & & & & \\
  $\beta_1$ & 1.000 & 0.999 & -0.001 & 0.280 & 0.268 & 0.995 & 0.278& 0.268 \\ 
  $\beta_2$ & -1.000 & -1.001 & -0.001 & 0.097 & 0.097 & -0.999 & 0.097&0.097  \\ 
  $\beta_3$ & 0.500 & 0.503 & 0.003 & 0.180 & 0.178 & 0.496 & 0.179& 0.178 \\ 
  $\theta$ & 0.500 & 0.499 & -0.001 &- & 0.393 & 0.400 & - &0.393 \\ 
      
  \hline
  parfm& & & & & & & & \\
  $\beta_1$ & 1.000 & 1.001 & 0.001 & 0.275 & 0.265 & 1.001 & 0.274 & 0.265\\ 
  $\beta_2$ & -1.000 & -1.004 & -0.004 & 0.095 & 0.096 & -1.005 & 0.094 &0.096 \\ 
  $\beta_3$ & 0.500 & 0.505 & 0.005 & 0.177 & 0.176 & 0.497 & 0.177 &0.176 \\ 
  $\theta$ & 0.500 & 0.465 & -0.035 & 0.221 & 0.388 & 0.353 & 0.180 &0.389 \\ 
  
  \hline
  frailtyEM& & & & &  & & & \\
  $\beta_1$ & 1.000 & 0.999 & -0.001 & 0.279 & 0.266 & 0.993 & 0.278 & 0.266 \\ 
  $\beta_2$ & -1.000 & -0.999 & 0.001 & 0.097 & 0.096 & -0.997 & 0.097 & 0.096\\ 
  $\beta_3$ & 0.500 & 0.503 & 0.003 & 0.179 & 0.174 & 0.496 & 0.178 &0.174 \\ 
  $\theta$ & 0.500 & 0.458 & -0.042 & 0.218 & 0.377 & 0.350 & 0.179 & 0.379\\ 
  
  \hline
   frailtySurv& & & & & & & & \\
   $\beta_1$ & 1.000 & 1.009 & 0.009 & 0.290 & 0.271 & 1.010 & 0.286 &0.271 \\ 
  $\beta_2$ & -1.000 & -1.005 & -0.005 & 0.103 & 0.101 & -1.001 & 0.101 & 0.101\\ 
  $\beta_3$ & 0.500 & 0.506 & 0.006 & 0.166 & 0.181 & 0.500 & 0.164 & 0.181\\ 
  $\theta$ & 0.500 & 0.465 & -0.035 & 0.226 & 0.375 & 0.365 & 0.151 & 0.376\\ 
   
   \hline
  frailtyHL& & & & & & & & \\
  $\beta_1$ & 1.000 & 1.001 & 0.001 & 0.279 & 0.269 & 0.999 & 0.278 &0.269 \\ 
  $\beta_2$ & -1.000 & -1.003 & -0.003 & 0.097 & 0.097 & -1.000 & 0.097 &0.097 \\ 
  $\beta_3$ & 0.500 & 0.504 & 0.004 & 0.179 & 0.178 & 0.497 & 0.178 &0.178 \\ 
  $\theta$ & 0.500 & 0.513 & 0.013 & 0.248 & 0.410 & 0.402 & 0.207 &0.410 \\ 
  
   \hline
   frailtypack& & & & & & & & \\
 $\beta_1$ & 1.000 & 1.023 & 0.023 & 0.272 & 0.275 & 1.023 & 0.274 &0.276 \\ 
  $\beta_2$ & -1.000 & -0.966 & 0.034 & 0.095 & 0.119 & -0.980 & 0.095 &0.120 \\ 
  $\beta_3$ & 0.500 & 0.499 & -0.001 & 0.178 & 0.177 & 0.491 & 0.177 &0.177 \\ 
  $\theta$ & 0.500 & 0.437 & -0.063 & 0.218 & 0.408 & 0.320 & 0.170 &0.412 \\ 
  
  \hline

\end{tabular}
\label{table2}
\end{table}

\newpage

\begin{table}[htb]
\tiny
\centering
\caption{Performance of the parameter estimation of different R packages. We only considered the converged fitted models for 1000 simulated datasets for each package. The total sample size is 400 and the censorship is 80\%.}
\begin{tabular}{lllllllll}
\hline
 Parameter &True &Mean&Bias& Mean.se  & Emp. se & Median&Median.se& MSE\\
\hline
 40 clusters of size 10 \\
  \hline
  survival& & & & & & & & \\
  $\beta_1$ &1&   0.985&	-0.015&	0.446&	0.478&	0.977&	0.442& 0.478\\ 
  $\beta_2$ &-1&  -0.993&	0.007&	0.146&	0.16&  -0.986&	0.145  & 0.160  \\
  $\beta_3$ & 0.5& 0.484&-0.016&	0.283&	0.286&	0.485&	0.281&0.286 \\
  $\theta$&0.5& 0.441&-0.059&	-	 &     0.276&	0.391&	- & 0.279 \\
 \hline
  parfm& & & & & & & & \\
  $\beta_1$ &1&1.006& 	0.006& 	0.442& 	0.474& 	0.995& 	0.439&0.474  \\
  $\beta_2$ &-1&-1.006& -0.006& 0.15& 	0.159& 	-1.001&	0.148& 0.159\\
  $\beta_3$ &0.5&0.487& -0.013& 0.284& 	0.285& 	0.494& 	0.282& 0.285\\
  $\theta$&0.5&0.467&  -0.033& 	0.243& 	0.277& 	0.428& 	0.231& 0.278 \\
  \hline
  frailtyEM& & & & &  & & & \\
  $\beta_1$ &1&0.994	&-0.006	&0.447	&0.469&	0.989&	0.443& 0.469\\
  $\beta_2$ & -1&-0.995	&0.005	&0.147	&0.156&  -0.986&    0.146&0.156 \\
  $\beta_3$ &0.5&0.484	&-0.016	&0.283	&0.280&	0.487&	0.281&0.280 \\
  $\theta$&0.5&0.446   &-0.054 &0.244	&0.273&	0.399&	0.232& 0.276\\
  \hline
   frailtySurv& & & & & & & & \\
  $\beta_1$ &1&0.976 &	-0.024&	0.572&	0.48&   0.98&	0.557&0.481 \\
  $\beta_2$ &-1&-0.998&	0.002&	0.161&	0.163&	-0.99	& 0.156&0.163 \\
  $\beta_3$ & 0.5&0.488& -0.012& 0.292&	0.29&	0.49& 0.285& 0.290\\
  $\theta$&0.5&0.436	&-0.064& 0.284&	0.284&	0.391& 0.26&0.288 \\
  \hline
  frailtyHL& & & & & & & & \\
  $\beta_1$ & 1&1.008&	0.008&	0.452&	0.488&	1.006&	0.448&0.488 \\
  $\beta_2$ & -1&-1.015& -0.015& 0.149&	0.165&	-1.007&	0.148&0.165 \\
  $\beta_3$ &0.5&0.495&	-0.005&	0.286&	0.293&	0.492&	0.284&0.293 \\
  $\theta$&0.5&0.525&	0.025&	0.261&	0.308&	0.481&	0.251& 0.309\\
  \hline
   frailtypack& & & & & & & & \\
  $\beta_1$ &1    &1.084& 	0.084&	0.425&	0.483&	1.081&	0.435& 0.490\\
  $\beta_2$ &-1   &-0.963&	0.037&	0.148&	0.172&	-0.953&	0.147&0.173 \\
  $\beta_3$ &0.5 &0.492& 	-0.008&	0.279&	0.282&	0.493&	0.278&0.282 \\
  $\theta$&0.5 &0.369&	-0.131&	0.21  &	0.249&	0.339&	0.21&0.266 \\
\hline
 10 clusters of size 40 \\
  \hline
survival& & & & & & & & \\
  $\beta_1$ & 1.000 & 0.997 & -0.003 & 0.457 & 0.448 & 0.981 & 0.452 &0.448 \\ 
  $\beta_2$ & -1.000 & -1.008 & -0.008 & 0.152 & 0.162 & -0.999 & 0.150& 0.162 \\ 
  $\beta_3$ & 0.500 & 0.491 & -0.009 & 0.288 & 0.293 & 0.488 & 0.285& 0.293 \\ 
  $\theta$ & 0.500 & 0.479 & -0.021 &- & 0.405 & 0.353 & - & 0.405\\ 
      
  \hline
  parfm& & & & & & & & \\
  $\beta_1$ & 1.000 & 1.008 & 0.008 & 0.446 & 0.432 & 1.000 & 0.443& 0.432 \\ 
  $\beta_2$ & -1.000 & -1.018 & -0.018 & 0.149 & 0.154 & -1.012 & 0.147 & 0.154\\ 
  $\beta_3$ & 0.500 & 0.498 & -0.002 & 0.284 & 0.289 & 0.492 & 0.281& 0.289  \\ 
  $\theta$ & 0.500 & 0.483 & -0.017 & 0.271 & 0.414 & 0.353 & 0.227& 0.414 \\ 
  
  \hline
  frailtyEM& & & & &  & & & \\
    $\beta_1$ & 1.000 & 0.998 & -0.002 & 0.455 & 0.428 & 0.980 & 0.450& 0.428 \\ 
  $\beta_2$ & -1.000 & -1.008 & -0.008 & 0.151 & 0.154 & -1.000 & 0.149& 0.154 \\ 
  $\beta_3$ & 0.500 & 0.493 & -0.007 & 0.286 & 0.283 & 0.497 & 0.284&0.283 \\ 
  $\theta$ & 0.500 & 0.472 & -0.028 & 0.269 & 0.395 & 0.351 & 0.228& 0.396 \\ 
  
  \hline
   frailtySurv& & & & & & & & \\
   $\beta_1$ & 1.000 & 1.008 & 0.008 & 0.554 & 0.451 & 0.992 & 0.526& 0.451 \\ 
  $\beta_2$ & -1.000 & -1.010 & -0.010 & 0.158 & 0.161 & -1.001 & 0.151& 0.161 \\ 
  $\beta_3$ & 0.500 & 0.497 & -0.003 & 0.275 & 0.295 & 0.494 & 0.267& 0.295 \\ 
  $\theta$ & 0.500 & 0.454 & -0.046 & 0.272 & 0.409 & 0.345 & 0.206& 0.411 \\ 
   
   \hline
  frailtyHL& & & & & & & & \\
  $\beta_1$ & 1.000 & 1.006 & 0.006 & 0.457 & 0.446 & 0.985 & 0.453& 0.446 \\ 
  $\beta_2$ & -1.000 & -1.015 & -0.015 & 0.152 & 0.161 & -1.007 & 0.150& 0.161 \\ 
  $\beta_3$ & 0.500 & 0.495 & -0.005 & 0.288 & 0.293 & 0.497 & 0.285&0.293  \\ 
  $\theta$ & 0.500 & 0.534 & 0.034 & 0.299 & 0.445 & 0.407 & 0.257&0.446  \\ 
  
   \hline
   frailtypack& & & & & & & & \\
 $\beta_1$ & 1.000 & 1.108 & 0.108 & 0.438 & 0.470 & 1.084 & 0.440& 0.481 \\ 
  $\beta_2$ & -1.000 & -0.980 & 0.020 & 0.150 & 0.170 & -0.972 & 0.149&0.170\\ 
  $\beta_3$ & 0.500 & 0.502 & 0.002 & 0.284 & 0.290 & 0.503 & 0.281&0.290\\ 
  $\theta$ & 0.500 & 0.411 & -0.089 & 0.244 & 0.398 & 0.304 & 0.207& 0.406 \\ 
  \hline
\end{tabular}
\label{table3}
\end{table}
\newpage

\begin{table}[htb]
\tiny
\centering
\caption{Coverage probability of the 95$\%$ CI for the estimated regression coefficients and \textcolor{red}{the variance of the frailty terms}. }

\begin{tabular}{lllllll}
\hline
Parameter &{ survival} &{  parfm}&{ frailtyEM} &{  frailtySurv}&{ frailtyHL}&{ frailtypack}\\
  \hline
  c = 20, 10 clusters of size 10& & & & & & \\
  $\beta_1$ &95.00  & 93.67 &94.86 & 95.29 & 95.04 & 88.53   \\ 
  $\beta_2$ &92.81 &92.11 &92.71 & 92.59 & 92.91 & 68.21   \\ 
  $\beta_3$ &94.40 &93.87 &94.32 & 91.38 & 94.53 & 94.77  \\ 
  CI$^{(1)}({\theta})$ &- &71.55 &69.99 & 67.94 & 77.20 & 48.50   \\ 
  CI$^{(2)}({\theta})$  &- &96.88 &87.34 & 91.80 & 96.73 & 99.40  \\ 
  
 \hline
  c = 20, 10 clusters of size 40 & & & & & &\\
  $\beta_1$ &95.8 &95.30 &96.05 & 93 & 95.87 & 95.47   \\ 
  $\beta_2$ &96 &94.99 &95.53 & 95 & 95.87 & 93.96   \\ 
  $\beta_3$ &95.3 &95.49 &95.32 & 91.3 & 95.26 & 95.17  \\ 
  CI$^{(1)}({\theta})$  &- &66.37 &64.76 & 67.7 & 72.68 & 65.06  \\ 
  CI$^{(2)}({\theta})$ &- &75.98 &77.68 & 72 & 81.15 & 74.82  \\ 

  \hline
   c= 20, 40 clusters of size 10& & & & & & \\
  $\beta_1$ &93.5 &94  &93.54 & 95.7 & 93 & 93.72   \\ 
  $\beta_2$ &92.9 &92.9 &92.73 & 93.4 & 93.1 & 89.56   \\ 
  $\beta_3$ &95.6 &95.3 &95.56 & 94.7 & 95.6 & 95.48  \\ 
  CI$^{(1)}({\theta})$ &- &70.2 &69.32 & 73.7 & 73.8 & 63.65  \\ 
  CI$^{(2)}({\theta})$  &- &82.6 &79.1 & 85.6 & 85 & 78.51  \\ 

   \hline
  c= 50, 10 clusters of size 10& & & & & & \\
  $\beta_1$ & 94.19 &92.77 &93.98 & 95.77 & 93.42 & 88.72   \\ 
  $\beta_2$ & 93.09 &94.12 &93.74 & 93.04 & 92.28 & 81.23    \\ 
  $\beta_3$ & 94.59 &94.69 &94.58 & 91.03 & 93.92 & 95.80   \\ 
  CI$^{(1)}({\theta})$ &- &80.37 &80.75 & 77.92 & 82.08 & 51.21\\ 
  CI$^{(2)}({\theta})$  &- & 97.06 &95.08 & 95.99 & 96.41 & 99.72\\ 
 \hline
  c = 50, 10 clusters of size 40 & & & & & & \\
  $\beta_1$ &95.8 &95.98 &95.96 & 93.1 & 95.57 & 94.47   \\ 
  $\beta_2$ &95.3 &95.07 &95.18 & 92.7 & 95.37 & 86.52   \\ 
  $\beta_3$ &94.5 &94.47 &95.18 & 89.7 & 94.47 & 94.16  \\ 
  CI$^{(1)}({\theta})$  &- &67.44 &65.03 & 60.2 & 73.24 & 62.78  \\ 
  CI$^{(2)}({\theta})$ &- &82.41 &80.65 & 72.3 & 86.59 & 81.29  \\ 
  \hline
 c= 50, 40 clusters of size 10& & & & & & \\
  $\beta_1$ &94.5 &94.7  &94.48 & 97.7 & 94.5 & 91.88   \\ 
  $\beta_2$ &94 &94.1     &94.18 & 95.3 & 93.9 & 85.26   \\ 
  $\beta_3$ &95.5 &95.8  &95.58 & 94.1 & 95.5 & 95.69  \\ 
  CI$^{(1)}({\theta})$  &- &75.6 &72.69 & 76.7 & 80 & 63.29  \\ 
  CI$^{(2)}({\theta})$ &- &92.1 &84.9 & 92.1 & 93.3 & 87.26  \\ 
 \hline
  c= 80, 10 clusters of size 10& & & & & & \\
  $\beta_1$ &94.96 &93.51 &94.58 & 97.35 & 94.53 & 82.24   \\ 
  $\beta_2$ &94.65  &94.87 &95.10 & 93.58 & 91.90 & 90.31   \\ 
  $\beta_3$ &94.24   &93.06 &93.53 & 92.76 & 92.67 & 92.73   \\ 
  CI$^{(1)}({\theta})$ &- & 98.64 &99.65 & 97.55 & 91.90 & 55.8  \\ 
  CI$^{(2)}({\theta})$  &- &90.50 &96.86 & 97.60 & 89.92 & 96.57  \\ 
 \hline
  c = 80, 10 clusters of size 40 & & & & & &\\
  $\beta_1$ &95.7 &95.42 &96.03 & 95.19 & 95.72 & 93.49   \\ 
  $\beta_2$ &93.8 &93.85 &93.99 & 91.88 & 93.78 & 89.98   \\ 
  $\beta_3$ &94.7 &93.54 &94.74 & 90.17 & 94.90 & 94.89  \\ 
  CI$^{(1)}({\theta})$  &- &77.08 &75.32 & 70.51 & 81.45 & 68.74  \\ 
  CI$^{(2)}({\theta})$  &- &93.96 &87.58 & 89 & 92.47 & 96.79  \\
  \hline
 c= 80, 40 clusters of size 10& & & & & & \\
  $\beta_1$ &92.7 &92.53 &92.90 & 97.4 & 92.96 & 91.37   \\ 
  $\beta_2$ &93.3 &93.86 &93.95 & 93.3 & 93.47 & 88.77   \\ 
  $\beta_3$ &94.9 &95.39 &94.89 & 94.9 & 94.57 & 95.29  \\ 
  CI$^{(1)}({\theta})$ &- &87.41 &86.01 & 87.6 & 90.95 & 76.33  \\ 
  CI$^{(2)}({\theta})$ &- &96.83 &93.04 & 97.8 & 93.25 & 98.60  \\ 
 
  \hline
  \end{tabular}
\label{cptable}
\end{table}

\newpage

\bibliography{ref}


\end{document}